\documentclass[aps,prd,preprint,groupedaddress]{revtex4-1}
\usepackage{graphicx}
\usepackage{amsmath,amssymb}
\usepackage{bm}

\begin{document}
\begin{flushright}
  OU-HET-777 \ \\
\end{flushright}
\vspace{0mm}


\title{Are there infinitely many decompositions
of the nucleon spin ?}


\author{M.~Wakamatsu}
\email[]{wakamatu@phys.sci.osaka-u.ac.jp}
\affiliation{Department of Physics, Faculty of Science, \\
Osaka University, \\
Toyonaka, Osaka 560-0043, Japan}



\begin{abstract}
We discuss the uniqueness or non-uniqueness problem of the decomposition
of the gluon field into the physical and pure-gauge components, which
is the basis of the recently proposed two physically inequivalent
gauge-invariant decompositions of the nucleon spin.
It is crucialy important to recognize the fact that the standard
gauge fixing procedure is essentially a process of projecting out the physical
components of the massless gauge field. A complexity of the nonabelian
gauge theory as compared with the abelian case is that a closed
expression for the physical component can be given only with use of
the non-local Wilson line, which is generally path-dependent.
It is known that, by choosing an infinitely long straight-line path in space
and time, the direction of which is characterized by a constant 4-vector $n^\mu$,
one can cover a class of gauge called the general axial gauge, containing
three popular gauges, i.e. the temporal, the light-cone, and the spatial axial gauge.
Within this general axial gauge, we have calculated the 1-loop evolution
matrix for the quark and gluon longitudinal spins in the nucleon.
We found that the final answer is exactly the same independently of
the choices of $n^\mu$, which amounts to proving the gauge-independence
and path-independence simultaneously, although within a restricted
class of gauges and paths.   
By drawing on all of these findings together with well-established
knowledge from gauge theories, we argue against the rapidly spreading view in
the community that there are infinitely many decompositions of the nucleon spin.
\end{abstract}


\maketitle


\section{Introduction}
Is a gauge-invariant complete decomposition of the nucleon spin possible ? 
It is a fundamentally important question of QCD as a color gauge theory.
The reason is that the gauge-invariance is generally believed to be a
{\it necessary condition} of {\it observability}.
Unfortunately, this is quite a delicate problem,
which is still under intense debate 
\cite{JM90}\nocite{Ji97PRL}\nocite{Ji97}\nocite{Ji98}\nocite{BJ99}\nocite{SW00}
\nocite{BLT04}\nocite{Chen08}\nocite{Chen08}\nocite{Chen11PRD}\nocite{Chen11PLB}
\nocite{Goldman11}\nocite{Tiwari08}\nocite{Ji11A}\nocite{Ji11B}\nocite{Wakamatsu10}
\nocite{Wakamatsu11A}\nocite{Wakamatsu11B}\nocite{Wakamatsu12}\nocite{BBC09}
\nocite{Leader11}\nocite{Leader12}\nocite{Cho10A}\nocite{Cho10B}\nocite{ZhangPak12}
\nocite{ZH11}\nocite{GS12}\nocite{Hatta11}\nocite{Hatta12}\nocite{Lorce12A}
\nocite{Lorce12B}\nocite{JXZ12}\nocite{JXY12A}\nocite{JXY12B}
\nocite{Lorce13}-\cite{GS13}.
In a series of papers \cite{Wakamatsu10}\nocite{Wakamatsu11A}
\nocite{Wakamatsu11B}-\cite{Wakamatsu12},
we have established the fact that there are two physically
inequivalent gauge-equivalent decompositions of the nucleon spin, which we
call the decompositions (I) and (II). The decompositions (I) and (II) are
respectively characterized by two different orbital angular momenta (OAMs)
for both of quarks and gluons, i.e. the ``dynamical" OAMs and the generalized
``canonical" OAMs. We also clarified the fact that difference of the above
two kinds of orbital angular momenta is characterized by a quantity which
we call the ``potential angular momentum", the QED correspondent of which
is nothing but the angular momentum carried by the electromagnetic field
or potential playing a key role in the famous Feynman paradox of classical
electrodynamics \cite{Wakamatsu10},\cite{FeynmanBook}.
The basic assumption for obtaining these two
gauge-invariant decompositions of the nucleon spin is that
the total gluon field can be decomposed into the two parts, i.e.
the physical component and the pure-gauge component, as 
$A^\mu (x) \, = \, A^\mu_{phys} (x) \, + \, A^\mu_{pure} (x)$.
In the course of deriving the above two gauge-invariant decompositions of
the nucleon spin, these two components are supposed to obey
the following general conditions, i.e. the pure-gauge condition
for the pure-gauge component,
$F^{\mu \nu}_{pure} \, \equiv \, \partial^\mu \,A^\nu_{pure} \, - \, 
\partial^\nu \,A^\mu_{pure} \, - \, i \,g \,
[\, A^\mu_{pure}, A^\nu_{pure} \,] \, = \, 0$,
supplemented with the homogeneous (or covariant) and inhomogeneous gauge
transformation properties respectively for the physical and pure-gauge
components of the gluon field under general gauge transformation of QCD. 

A natural question is whether these general conditions are enough to
uniquely fix the above decomposition.
The answer is evidently No !
Note however that the above decomposition is proposed as a
covariant generalization of Chen et al.'s decomposition given in a
noncovariant form as
$\mbox{\boldmath $A$} (x) \, = \, \mbox{\boldmath $A$}_{phys} (x) \, + \, 
\mbox{\boldmath $A$}_{pure} (x)$ \cite{Chen08},\cite{Chen09}.
One must know the fact that,
at least in the QED case, this decomposition is nothing
new. It just corresponds to the standardly-known transverse-longitudinal
decomposition of the 3-vector potential of the photon field, i.e. 
$\mbox{\boldmath $A$} (x) \, = \, \mbox{\boldmath $A$}_\perp (x) \, + \, 
\mbox{\boldmath $A$}_\parallel (x)$ satisfying the properties 
$\nabla \cdot \mbox{\boldmath $A$}_\perp \, = \, 0$ and 
$\nabla \times \mbox{\boldmath $A$}_\parallel \, = \, 0$ \cite{BookBLP82},
\cite{BookCDG89}.
It is a well-established fact that this decomposition is {\it unique}
once the Lorentz frame of reference is
specified \cite{BookCDG89}.
As we shall see later, a physically essential element here is the
transversality condition $\nabla \cdot \mbox{\boldmath $A$}_\perp = 0$
for the transverse (or physical) component of $\bm{A}$ given in a non-covariant form.
Naturally, a certain substitute
of this condition is necessary to uniquely fix the physical component of
$A^\mu_{phys}$ in
the above-mentioned decomposition given in a (seemingly) covariant form.
This fundamental fact of gauge theory is missed out in the
community, and conflicting views have rapidly spread around.

On the one hand, Lorc\'{e} claims that the above decomposition
is not unique because of the presence of what-he-call the
Stueckelberg symmetry, which alters both of $A^\mu_{phys}$ and
$A^\mu_{pure}$ while keeping their sum unchanged \cite{Lorce12A},\cite{Lorce12B}.
This misapprehension comes from the oversight of the importance of the
transversality condition that should be imposed on the physical component.
On the other, another argument against the uniqueness of the
above-mentioned decomposition is advocated by Ji et al.
\cite{JXZ12}\nocite{JXY12A}-\cite{JXY12B}.
According to them, the Chen
decomposition is a gauge-invariant extension (GIE) of the Jaffe-Manohar
decomposition based on the Coulomb gauge, while the Bashinsky-Jaffe
decomposition is a GIE of the Jaffe-Manohar decomposition based on the
light-cone gauge.
They claim that, since the way of GIE with use of path-dependent Wilson line
is not unique at all, there is no need that the above
two decompositions give the same physical predictions.
This made Ji reopen his longstanding claim that the gluon spin
$\Delta G$ in the nucleon is not a gauge-invariant quantity in a {\it true} or 
{\it traditional} sense, although it is a measurable quantity in
polarized deep-inelastic scatterings \cite{JTH96},\cite{HJL98}.
One should recognize a self-contradiction inherent in this claim.
In fact, first remember the fundamental proposition of physics, which states
that ``Observables must be gauge-invariant."
(Note that we are using the word ``observables'' in a strict sense.
That is, they must be quantities, which can be extracted purely experimentally,
i.e. without recourse to particular theoretical schemes or models.) 
The contraposition of this proposition (note that it is always correct if the
original proposition is correct) is gGauge-variant quantities cannot be observables".
This dictates that, if $\Delta G$ is claimed to be observable, it must
be gauge-invariant also in the {\it traditional} sense.

In view of the above-explained frustrated status, we believe it urgent
to correct widespread misunderstanding on the meaning of {\it true} or
{\it traditional} gauge-invariance in the problem of nucleon spin decomposition.
The paper is then organized as follows. In sect.II, we first clarify the fact
that, at least in the case of abelian gauge theory,
the decomposition of the gauge field into the physical and pure-gauge
component is nothing but the well-known transverse-longitudinal decomposition
of the vector potential.
It is a well-established fact that this decomposition is unique
as far as we are working in a prescribed Lorentz frame. 
We also point out a hidden problem of the gauge-invariant extension approach,
i.e. the {\it path-dependence}, through a concise pedagogical review of the
gauge-invariant formulation of the electromagnetism with use of the
nonlocal gauge link.
Next in sect.III, we give an explicit form of the physical component of
the gluon field based on a geometrical formulation of the nonabelian
gauge theory, which also uses a path-dependent Wilson line.
After clarifying an inseparable connection between the choice of path
contained in the Wilson line and the choice of gauge, we consider a special
class of paths, i.e. infinitely long straight-line paths, the
direction of which is characterized by a constant 4-vector $n^\mu$.
This particular choice of path is known to be equivalent to taking the
so-called general axial gauge, which contains in it three popular
gauges, i.e. the temporal, the light-cone, and the spatial axial gauges. 
Based on this general axial gauge specified by the 4-vector $n^\mu$,
we shall calculate the 1-loop evolution matrix for the quark and gluon
longitudinal gluon spins in the nucleon, in order to check whether
the answer depends on the choice of $n^\mu$, which characterizes simultaneously
the gauge choices within the general axial gauge and
the direction of the straight-line path in the geometric formulation.
Concluding remarks will then be given in sect.V.

\section{Critiques on the idea of St\"{u}ckelberg symmetry and
gauge-invariant-extension approach}

In a series of papers \cite{Wakamatsu10}\nocite{Wakamatsu11A}\nocite{Wakamatsu11B}
-\cite{Wakamatsu12}, we have shown that there are two physically
inequivalent decompositions of the nucleon spin, which we call
the decomposition (I) and (II).
The QCD angular momentum tensor in the decomposition (I) is given as
follows :
\begin{equation}
 M^{\mu \nu \lambda} \ = \ M^{\mu \nu \lambda}_{q-spin} \ + \ 
 M^{\mu \nu \lambda}_{q-OAM} \ + \ M^{\mu \nu \lambda}_{G-spin}
 \ + \ M^{\mu \nu \lambda}_{G-OAM} \ + \ M^{\mu \nu \lambda}_{boost},
 \label{decompositionI}
\end{equation}
with
\begin{eqnarray}
 M^{\mu \nu \lambda}_{q-spin}
 &=& \frac{1}{2} \,\epsilon^{\mu \nu \lambda \sigma} \,\bar{\psi} 
 \,\gamma_{\sigma} \,\gamma_5 \,\psi , \label{decompositionIA} \\
 M^{\mu \nu \lambda}_{q-OAM}
 &=& \bar{\psi} \,\gamma^\mu \,(\,x^{\nu} \,i \,D^{\lambda} 
 \ - \ x^{\lambda} \,i \,D^{\nu} \,) \,\psi \label{decompositionIB} \\
 M^{\mu \nu \lambda}_{G-spin} 
 &=& 2 \,\mbox{Tr} \,[\, F^{\mu \lambda} \,A^{\nu}_{phys} 
 \ - \ F^{\mu \nu} \,A^{\lambda}_{phys} \,], \label{decompositionIC} \\
 M^{\mu \nu \lambda}_{G-OAM} 
 &=& - \,2 \,\mbox{Tr} \,[\, F^{\mu \alpha} \,
 (\,x^{\nu} \,D^{\lambda}_{pure}
 \ - \ x^{\lambda} \,D^{\nu}_{pure} \,) \,A_{\alpha}^{phys} \,],
 \nonumber \\
 &\,& + \, 2 \,\mbox{Tr} \,[\, (\,D_{\alpha} \,F^{\alpha \mu} \,)
 \,(\,x^{\nu} \,A^{\lambda}_{phys} \ - \ 
 x^{\lambda} \,A^{\nu}_{phys} \,) \,], \label{decompositionID}
\end{eqnarray}
and
\begin{equation}
 M^{\prime \mu \nu \lambda}_{boost} 
 \ = \ - \,\frac{1}{2} \,\mbox{Tr} \,F^2 \,(\,x^{\nu} \,g^{\mu \lambda} 
 \ - \ x^{\lambda} \,g^{\mu \nu} \,) . \label{decompositionIE}
\end{equation}

On the other hand, the QCD angular momentum tensor in the decomposition (II)
is given as follows :
\begin{equation}
 M^{\mu \nu \lambda} \ = \ M^{\prime \mu \nu \lambda}_{q-spin} \ + \ 
 M^{\prime \mu \nu \lambda}_{q-OAM} \ + \ M^{\prime \mu \nu \lambda}_{G-spin}
 \ + \ M^{\prime \mu \nu \lambda}_{G-OAM} \ + \ M^{\prime \mu \nu \lambda}_{boost},
 \label{decompositionII}
\end{equation}
with
\begin{eqnarray}
 M^{\prime \mu \nu \lambda}_{q-spin}
 &=& M^{\mu \nu \lambda}_{q-spin}, \\
 M^{\prime \mu \nu \lambda}_{q-OAM}
 &=& \bar{\psi} \,\gamma^\mu \,(\,x^{\nu} \,i \,D^{\lambda}_{pure} 
 \ - \ x^{\lambda} \,i \,D^{\nu}_{pure} \,) \,\psi \label{decomposition2B} \\
 M^{\prime \mu \nu \lambda}_{G-spin} 
 &=& M^{\mu \nu \lambda}_{G-spin}, \label{decompositionIIC} \\
 M^{\prime \mu \nu \lambda}_{G-OAM} 
 &=& - \,2 \,\mbox{Tr} \,[\, F^{\mu \alpha} \,
 (\,x^{\nu} \,D^{\lambda}_{pure}
 \ - \ x^{\lambda} \,D^{\nu}_{pure} \,) \,A_{\alpha}^{phys} \,] ,
 \label{decompositionIID} \\
 M^{\prime \mu \nu \lambda}_{boost} &=& M^{\mu \nu \lambda} . 
 \label{decompositionIIE}
\end{eqnarray}
In these two decompositions, the quark and gluon intrinsic spin
parts are just common, and the difference lies only in the orbital parts.
The difference is given as follows : 
\begin{eqnarray}
 M^{\mu \nu \lambda}_{q-OAM} - M^{\prime \mu \nu \lambda}_{q-OAM}
 &=& \ - \,\left(\,
 M^{\mu \nu \lambda}_{G-OAM} - M^{\prime \mu \nu \lambda}_{G-OAM}
 \,\right) \nonumber \\
 &=& \ 2 \,\mbox{Tr} \,[\, (\,D_{\alpha} \,F^{\alpha \mu} \,)
 \,(\,x^{\nu} \,A^{\lambda}_{phys} \ - \ 
 x^{\lambda} \,A^{\nu}_{phys} \,) \,] .
\end{eqnarray}
The quantity characterizing the difference between the two kinds of orbital
angular momenta of quarks and gluons, i.e.
the quantity appearing in the r.h.s. of the above relation, is a
covariant generalization of the following quantity
\begin{equation}
 \bm{L}_{pot} \ = \ \int \rho^a \,(\bm{r} \times \bm{A}^a) \,d^3 r
\end{equation}
which we called the {\it potential angular momentum} in \cite{Wakamatsu10}.
The reason is that this just corresponds to the angular momentum carried by the
electromagnetic field or potential appearing in famous
Feynman's paradox of classical electrodynamics \cite{FeynmanBook}.
(For an interesting phenomenological implication concerning the difference
between these two physically inequivalent decompositions of the nucleon spin,
we refer to the references \cite{Waka10}\nocite{MT88}\nocite{Thomas08}
\nocite{WT05}\nocite{WN06}-\cite{WN08}.)

The whole argument above is based on the decomposition of the gluon field
$A^\mu$ into the physical component and the pure-gauge component as
\begin{equation}
 A^\mu \ = \ A^\mu_{phys} \ + \ A^\mu_{pure},
\end{equation}
satisfying the following general conditions, i.e. the pure-gauge condition
for $A^\mu_{pure}$
\begin{equation}
 F^{\mu \nu}_{pure} \ \equiv \ \partial^\mu A^\nu_{pure} \ - \ 
 \partial^\nu A_{pure} \ - \ i \,g \,[\,A^\mu_{pure}, A^\nu_{pure}] \ = \ 0,
\end{equation}
supplemented with the gauge-transformation properties for $A^\mu_{phys}$
and $A^\mu_{pure}$
\begin{eqnarray}
 A^\mu_{phys} (x) \ &\rightarrow& \ U(x) \,A^\mu_{phys} (x) \,U^\dagger (x), 
 \label{Eq:Tr_homogeneous} \\
 A^\mu_{pure} (x) \ &\rightarrow& \ U(x) \,\left( A^\mu_{pure} (x) \ + \ 
 \frac{i}{g} \,\,\partial^\mu \right) \,U^\dagger (x), \label{Eq:Tr_inhomogeneous}
\end{eqnarray}
under an arbitrary gauge transformation $U(x)$ of QCD.

In recent papers  \cite{Lorce12A},\cite{Lorce12B}, Lorce criticized that
the pure-gauge condition
$F^{\mu \nu}_{pure} = 0$ is insufficient to uniquely determine the
decomposition $A^\mu = A^\mu_{phys} + A^\mu_{pure}$. According to him,
there exists a hidden symmetry, which he calls a Stueckelberg symmetry.
In the simpler case of abelian gauge theory, the
proposed Stueckelberg transformation is given by
\begin{eqnarray}
 A^\mu_{phys} (x) \ &\rightarrow& \ A^\mu_{phys,g} (x) \ = \ 
 A^\mu_{phys} (x) \ - \ \partial^\mu C(x), \\
 A^\mu_{pure} (x) \ &\rightarrow& \ A^\mu_{pure,g} (x) \ = \ 
 A^\mu_{pure} (x) \ + \ \partial^\mu C(x), 
\end{eqnarray}
with $C (x)$ being an arbitrary function of space and time.
Certainly, this transformation changes both of $A^\mu_{phys}$ and 
$A^\mu_{pure}$, but the sum of them is intact.
It was then claimed that this hidden symmetry dictates the existence of
infinitely many decompositions of the gauge field into the physical and
pure-gauge components, thereby leading him to the conclusion that there are
in principle {\it infinitely many decompositions} of the nucleon spin. 

It is certainly true that the pure-gauge condition, together with the
homogeneous and inhomogeneous transformation properties of
$A^\mu_{phys}$ and $A^\mu_{pure}$, are not sufficient to
determine the decomposition $A^\mu = A^\mu_{phys} + A^\mu_{pure}$ uniquely.
However, one should remember the original motivation of this decomposition.
In the QED case with noncovariant treatment by
Chen et al. \cite{Chen08},\cite{Chen09}, this decomposition
is nothing more than the standard decomposition of the vector potential $\bm{A}$
of the photon field into the transverse and longitudinal components :
\begin{equation}
 \bm{A} (x) \ = \ \bm{A}_\perp (x) \ + \ \bm{A}_\parallel (x),
\end{equation}
where the transverse component and the longitudinal component are respectively
required to obey divergence-free and irrotational
conditions \cite{BookBLP82},\cite{BookCDG89} : 
\begin{equation}
 \nabla \cdot \bm{A}_\perp \ = \ 0, \ \ \ \ \ 
 \nabla \times \bm{A}_\parallel \ = \ 0.
\end{equation}
For the sake of later discussion, we also recall the fact that the
transverse-longitudinal decomposition can be made explicit with use of the
corresponding projection operators as follows : 

\begin{equation}
 A^i (x) \ = \ A^i_\perp (x) \ + \ A^i_\parallel (x) \ = \ 
 (\,P^{i j}_T \ + \ P^{i j}_L \,) \,A^j (x) ,
\end{equation}
with
\begin{eqnarray}
 P^{i j}_T \ &=& \ \delta^{i j} \ - \ \frac{\nabla^i \,\nabla^j}{\nabla^2}, \\
 P^{i j}_L \ &=& \ \ \frac{\nabla^i \,\nabla^j}{\nabla^2} .
\end{eqnarray}
As is well-known, these two components transform as follows,
\begin{eqnarray}
 \bm{A}_\perp (x) \ &\rightarrow& \ \bm{A}^\prime_\perp (x) \ = \ \bm{A}_\perp (x),
 \label{Eq:gauge_trans_T} \\
 \bm{A}_\parallel (x) \ &\rightarrow& \ \bm{A}^\prime_\parallel (x) \ = \ 
 \bm{A}_\parallel (x) \ - \ \nabla \,\Lambda (x). \label{Eq:gauge_trans_L} 
\end{eqnarray}
under a general abelian gauge transformation. 
This means that $\bm{A}_\parallel$
carries unphysical gauge degrees of freedom, while $\bm{A}_\perp$ is
absolutely intact under an arbitrary gauge transformation.
Besides, it is a well-established fact that this decomposition is {\it unique},
once the Lorentz-frame of reference is specified \cite{BookCDG89}. 
(To be more precise,
the uniqueness is guaranteed by a supplemental condition that $\bm{A}$
falls off faster than $1 / r^2$ at the spatial infinity, which is
satisfied in usual circumstances that happen in the electromagnetism.)
This uniqueness of the decomposition indicates that, in QED, there exists no
Stueckelberg symmetry as suggested by Lorc\'{e}. In fact, within the
above-mentioned noncovariant framework, the Stueckelberg transformation
a la Lorc\'{e} reduces to
\begin{eqnarray}
 \bm{A}_\perp (x) \ &\rightarrow& \ \bm{A}^g_\perp (x) \ = \ 
 \bm{A}_\perp (x) \ + \ \nabla \,C(x), \\
 \bm{A}_\parallel (x) \ &\rightarrow& \ \bm{A}^g_\parallel (x) \ = \ 
 \bm{A}_\parallel (x) \ - \ \nabla \,C(x).
\end{eqnarray}
One can convince that the transformed longitudinal component
$\bm{A}^g_\parallel (x)$ retains the irrotational property,
\begin{equation}
 \nabla \times \bm{A}^g_\parallel \ = \ \nabla \times 
 (\bm{A}_\parallel \ - \ \nabla \,C(x) ) \ = \ \nabla \times \bm{A}_\parallel
 \ = \ 0.
\end{equation}
(This is simply a reflection of the fact the standard gauge transformation for
$\bm{A}_\parallel$ keeps the magnetic field $\bm{B} = \nabla \times \bm{A}$ intact.)
However, one finds that the transformed component $\bm{A}^g_\perp (x)$ does
not satisfy the desired divergence-free (or transversality) condition
$\nabla \cdot \bm{A}^g_\perp = 0$ any more, since
\begin{equation}
 \nabla \cdot \bm{A}^g_\perp \ = \ \nabla \cdot 
 (\bm{A}_\perp \ + \ \nabla \,C (x) ) \ = \ 
 \Delta \,C(x) \ \neq \ 0,
\end{equation}
unless $\Delta C(x) = 0$. (As a matter of course, different from the
Stueckelberg transformation, there is no such problem in the
standard gauge transformation (\ref{Eq:gauge_trans_T}) and 
(\ref{Eq:gauge_trans_L}), because $\bm{A}_\perp$ is
intact under a general gauge transformation.)
The condition $\Delta C(x) = 0$ means that $C(x)$
is a harmonic function in three spatial dimension. If it is required to
vanish at the spatial infinity, it must be identically zero owing to
the Helmholtz theorem.
As is clear from the discussion above, the Stueckelberg-like transformation
does not generally preserve the transversality condition of the transverse or
physical component of $\bm{A}$.
In other words, the Stueckelberg symmetry does not actually exist
and/or it has nothing to do with a physical symmetry of QED.
Let us repeat again the well-founded fact in QED.
The transverse-longitudinal decomposition is unique once the
Lorentz-frame of reference is fixed. 

Still, a bothersome problem here is that the transverse-longitudinal
decomposition is not a relativistically invariant manipulation.
A vector field that
appears transverse in a certain Lorentz frame is not necessarily
transverse in  another Lorentz frame.
An immediate question is then what meaning one can give to the seemingly
covariant decomposition of the gauge field like
$A^\mu = A^\mu_{phys} + A^\mu_{pure}$.
Putting it in another way, in view of the fact that the
transverse-longitudinal decomposition
can be made only at the sacrifice of breaking the Lorentz-covariance, how can
we get an explicit form of this decomposition, which is usable in a desired
Lorentz frame ? Leaving this nontrivial question aside, we want to make
some general remarks on the treatment of gauge theories. 
In a covariant treatment of gauge theories, we start with
the gauge field $A^\mu$ with four components ($\mu = 0, 1, 2, 3$).
We however know that
the massless gauge field has only two independent dynamical degrees of freedom,
i.e., two transverse components, say $A^1$ and $A^2$. The other two components,
i.e. the scalar component $A^0$ and the longitudinal component $A^3$,
are not independent dynamical degrees of freedom.
For quantizing a gauge theory, we need a procedure of gauge-fixing.
A gauge-fixing procedure is essentially an operation, which eliminates
the unphysical degrees of freedom so as to pick out the two transverse components.
In this sense, the transverse-longitudinal decomposition and
the gauge-fixing procedure are closely interrelated operations (one might
say that they are almost {\it synonymous}), even though
they are not absolutely identical operation.

Another argument against the uniqueness of the nucleon spin decomposition is
based on the idea of gauge-invariant extension with use of path-dependent
Wilson line \cite{JXZ12}\nocite{JXY12A}-\cite{JXY12B}.
The idea of gauge-link in gauge theories is of more general
concerns and has a long history.
Once, DeWitt tried to formulate the quantum electrodynamics in a
gauge-invariant way, i.e. without introducing gauge-dependent
potential \cite{DeWitt62}.
However, it was recognized soon that, although the framework
is manifestly gauge-invariant by construction it {\it does} depend of the choice
of path defining the gauge-invariant potential \cite{Belinfante62}
\nocite{Mandelstam62}\nocite{RohrlichStrocchi65}-\cite{Yang85}.
Since the problem seems to be intimately connected with the one we are
confronting with, we think it instructive to briefly review this framework
by paying attention to its delicate point.

According to DeWitt, once given an appropriate set of electron and photon fields
$(\psi (x), \,A_\mu (x))$, the gauge-invariant set of
the electron and photon fields $(\psi^\prime (x), \,A^\prime_\mu (x)$) can
be constructed as
\begin{eqnarray}
 \psi^\prime (x) \ &\equiv& \ \ 
 e^{\,i \,e \,\Lambda (x)} \,\,\psi (x), \label{GI_electron} \\
 A^\prime_\mu (x) \ &\equiv& \ 
 A_\mu (x) \ + \ \partial_\mu \Lambda (x) . \label{GI_photon}
\end{eqnarray}
with 
\begin{equation}
 \Lambda (x) \ = \ - \,\int_{- \,\infty}^0 \,A_\sigma (z) \,\,
 \frac{\partial z^\sigma}{\partial \xi} \,\,d \xi, \label{gauge_tr_func}
\end{equation}
where $z^\mu (x,\xi)$ is a point on the line toward $x$, with $\xi$ being
a parameter chosen in such a way that
\begin{equation}
 z^\mu (x,0) \ = \ x^\mu, \ \ \ \mbox{and} \ \ \ 
 z^\mu (x, - \,\infty) \ = \ \mbox{spatial infinity}.
\end{equation}
Note here that $\partial z^\mu \,/ \, \partial x^\lambda = \delta_\lambda{}^\mu$
at $\xi = 0$.

One can easily convince that the new electron and photon fields defined by
(\ref{GI_electron}) and (\ref{GI_photon}) are in fact gauge-invariant.
In fact, under an arbitrary gauge transformation
\begin{eqnarray}
 \psi (x) \ &\rightarrow& \  
 e^{\,i \,e \,\omega (x)} \,\psi (x), \\
 A_\mu (x) \ &\rightarrow& \  
 A_\mu (x) \ + \ \partial_\mu \omega (x),
\end{eqnarray}
the function $\Lambda (x)$ transforms as
\begin{eqnarray}
 \Lambda (x) \ &\rightarrow& \ - \,\int_{- \,\infty}^0 \,
 \left( A_\sigma (z) \ + \ \partial_\sigma \omega (z) \right) \,
 \frac{\partial z^\sigma}{\partial \xi} \,d \xi \nonumber \\
 &=& \ - \,\int_{- \,\infty}^0 \,A_\sigma (z) \,
 \frac{\partial z^\sigma}{\partial \xi} \,d \xi \ - \
 \int_{- \,\infty}^0 \,\frac{\partial \omega (z)}{\partial \xi} \,d \xi
 \ = \ 
 \Lambda (x) \ - \ \omega (x). 
\end{eqnarray}
This means that $\psi^\prime (x)$ transforms as
\begin{eqnarray}
 \psi^\prime (x) \ &\rightarrow& \ e^{\,i \,e \,(\Lambda (x) - \omega (x))} \,\,
 e^{\,i \,e \,\omega (x)} \,\,\psi (x) \nonumber \\
 &=& \ e^{\,i \,e \,\Lambda (x)} \,\,\psi (x) \ = \ 
 \psi^\prime (x),
\end{eqnarray}
that is, $\psi^\prime (x)$ is gauge-invariant.
The gauge-invariance of $A^\prime_\mu (x)$ can also be easily verified.
For instructive purpose, we reproduce here the proof.
The manipulation goes as follows :
\begin{eqnarray}
 A^\prime_\mu (x) \ &=& \ A_\mu (x) \ + \ \partial_\mu \Lambda (x) \nonumber \\
 &=& \ A_\mu \ - \ \partial_\mu \,\int_{- \,\infty}^0 \,
 A_\sigma (z) \,\frac{\partial z^\sigma}{\partial \xi} \,d \xi \nonumber \\
 &=& \ A_\mu \ - \ \int_{- \,\infty}^0 \,\partial_\nu A_\sigma (z) \,
 \frac{\partial z^\nu}{\partial x^\mu} \,\frac{\partial z^\sigma}{\partial \xi}
 \,d \xi \ - \ 
 \int_{- \,\infty}^0 \,A_\sigma (z) \,\frac{\partial}{\partial \xi} \,
 \left( \frac{\partial z^\sigma}{\partial x^\mu} \right) \,d \xi \nonumber \\
 &=& \ A_\mu \ - \ \int_{- \,\infty}^0 \,\partial_\nu A_\sigma (z) \,
 \frac{\partial z^\nu}{\partial x^\mu} \,
 \frac{\partial z^\sigma}{\partial \xi} \,d \xi \nonumber \\
 &\,& \hspace{8mm} + \   
 \int_{- \,\infty}^0 \,\partial_\nu \,A_\sigma (z) \,
 \frac{\partial z^\nu}{\partial \xi} \,\frac{\partial z^\sigma}{\partial x^\mu}
 \,d \xi \ - \ A_\sigma (z) \,
 \left. \frac{\partial z^\sigma}{\partial x^\mu} \right|^{\xi = 0}_{\xi = - \,\infty}
 \nonumber \\ 
 &=& \ A_\mu \ - \ \int_{- \,\infty}^0 \,\partial_\nu A_\sigma (z) \,
 \frac{\partial z^\nu}{\partial x^\mu} \,
 \frac{\partial z^\sigma}{\partial \xi} \,d \xi \nonumber \\
 &\,& \hspace{8mm} + \   
 \int_{- \,\infty}^0 \,\partial_\nu \,A_\sigma (z) \,
 \frac{\partial z^\nu}{\partial \xi} \,\frac{\partial z^\sigma}{\partial x^\mu}
 \,d \xi \ - \ A_\sigma (x) \,\delta_\mu{}^\sigma \nonumber \\
 &=& \ - \,\int_{- \,\infty}^0 \,
 (\,\partial_\nu \,A_\sigma \ - \ \partial_\sigma \,A_\nu \,) \,
 \frac{\partial z^\nu}{\partial x^\mu} \,
 \frac{\partial z^\sigma}{\partial \xi} \,d \xi .
\end{eqnarray}
We thus find the key relation
\begin{eqnarray}
 A^\prime_\mu (x) \ &=& \ 
 - \,\int_{- \,\infty}^0 \,F_{\nu \sigma} (z) \,\,
 \frac{\partial z^\nu}{\partial x^\mu} \,
 \frac{\partial z^\sigma}{\partial \xi} \,d \xi. \label{GI_potential}
\end{eqnarray}
Since the r.h.s. of the above relation is expressed only in terms of gauge-invariant
field-strength tensor, the gauge-invariance of $A^\prime_\mu (x)$ is obvious.
This is the essence of the gauge-invariant formulation of QED by DeWitt.
Here is a catch, however. Although the r.h.s. of (\ref{GI_potential})
is certainly gauge-invariant, it generally depends on the path connecting
the point $x$ and spatial infinity.
To see it most transparently, let us take constant-time paths in a given Lorentz
frame, with the property $\partial z^0 \,/\, \partial \xi = 0$.
In this case, Eq. (\ref{gauge_tr_func}) reduces to
\begin{equation}
 \Lambda (x) \ = \ - \,\int_{- \,\infty}^x \,\bm{A} (x^0, \bm{z}) \cdot d \bm{z}.
\end{equation}
Let us now consider two space-like (or constant-time) paths $L_1$ and $L_2$
connecting $x$ and spatial infinity \cite{Belinfante62}.
The corresponding gauge-invariant electron fields are given by
\begin{eqnarray}
 \psi^\prime (x \,; \,L_1) \ &=& \ \exp \,\left[\, - i \, e \,
 \int_{L_1}^x \,\bm{A} (x^0, \bm{z}) \cdot d \bm{z} \,\right] \,\psi (x), \\
 \psi^\prime (x \,; \,L_2) \ &=& \ \exp \,\left[\, - i \, e \,
 \int_{L_2}^x \,\bm{A} (x^0, \bm{z}) \cdot d \bm{z}
 \,\right] \,\psi (x). 
\end{eqnarray}
These two gauge-invariant electron fields are related through
\begin{equation}
 \psi^\prime (x \,; \,L_1) \ = \ 
 \exp \,\left[\, i \,e \,\left(\,\int_{L_1}^x \ - \ 
 \int_{L_2}^x \,\right) \,\bm{A} (x^0, \bm{z}) \cdot d \bm{z} 
 \,\right] \,\psi^\prime (x ; L_2).
\end{equation}
Closing the path of integration to a loop $L$ by a connection at spatial infinity,
where all fields and potentials are assumed to vanish, we obtain
\begin{eqnarray}
 \psi^\prime (x \,; \,L_1) \ &=& \ \exp \,\left[\, i \,e \,
 \oint_L \,\bm{A} (x^0, \bm{z}) \cdot d \bm{z} \,\right] \,
 \psi^\prime (x \,; \,L_2)
 \nonumber \\
 &=& \ \exp \,\left[\, i \,e \, \iint_S \,
 ( \nabla_z \times \bm{A} (x^0, \bm{z})) \cdot d \bm{z} \,\right] \,
 \psi^\prime (x \,; \,L_2) \nonumber \\
 &=& \ \exp \,\left[\, i \,e \, \iint_S \,
 \bm{B} (x^0, \bm{z})) \cdot d \bm{z} \,\right] \, \psi^\prime (x \,; \,L_2). 
\end{eqnarray}
Since the magnetic flux does not vanish in general, $\psi^\prime (x \,; \,L_1)$
and $\psi^\prime (x \,; \,L_2)$ do not coincide, which means that
$\psi^\prime (x)$ is generally {\it path-dependent}.

Very interestingly, there is one interesting choice of $\Lambda (x)$,
which enables us to construct $\psi^\prime (x)$ and $A^\prime_\mu (x)$, which
are path-independent as well as gauge-invariant \cite{KT94},\cite{KT97}.
The choice corresponds to taking as
\begin{equation}
 \Lambda (x) \ = \ - \,\int_{- \,\infty}^x \,\bm{A}_\parallel (x^0, \bm{z}) 
 \cdot d \bm{z},
\end{equation}
where $\bm{A}_\parallel (x)$ is the longitudinal component in the decomposition
$\bm{A} (x) \ = \ \bm{A}_\perp (x) \ + \ \bm{A}_\parallel (x)$,
with the important properties
$\nabla \cdot \bm{A}_\perp \ = \ 0, \  
 \nabla \times \bm{A}_\parallel \ = \ 0$.
Interestingly, since
\begin{equation}
 \oint_L \,\bm{A}_\parallel (x^0, \bm{z}) \cdot d \bm{z} \ = \ 
 \iint_S \,\left( \nabla_z \times \bm{A}_\parallel (x^0, \bm{z}) \right) 
 \cdot d \bm{S} \ = \ 0,
\end{equation}
due to the irrotational property of $\bm{A}_\parallel (x)$,
the electron wave function defined by
\begin{equation}
 \psi^\prime (x) \ = \ \exp \,\left[\, - i \,\,e \,
 \int_{- \,\infty}^x \,\bm{A}_\parallel (x^0, \bm{z}) \cdot d \bm{z}
 \,\right] \,\psi (x)
\end{equation}
is not only gauge-invariant but also path-independent.
We also recall the fact that the transverse and longitudinal components of
$\bm{A}$ can be expressed as
\begin{equation}
 \bm{A}_\perp (x) \ = \ \bm{A} (x) \ - \ \nabla \,\frac{1}{\nabla^2} \,
 \nabla \cdot \bm{A} (x), \ \ \ \ 
 \bm{A}_\parallel (x) \ = \ \nabla \,\frac{1}{\nabla^2} \,\nabla \cdot \bm{A} (x).
\end{equation}
Therefore, $\psi^\prime (x)$ can be reduced to the following form,
\begin{eqnarray}
 \psi^\prime (x) \ &=& \ \exp \,\left[\, - \,i \,e \,
 \int_{- \,\infty}^x \,\left(\, \nabla_z \,\frac{1}{\nabla^2_z} \,
 \nabla_z \cdot \bm{A} (x^0, \bm{z}) \,\right) \cdot d \bm{z} \,\right] \,
 \psi (x) \nonumber \\
 &=& \exp \,\left[\, - \,i \,e \,\frac{\nabla \cdot \bm{A}}{\nabla^2} \,(x) 
 \,\right] \,\psi (x) .
\end{eqnarray}
Note that, in this form, the path-independence of $\psi^\prime (x)$ is
self-evident. We recall that this quantity is nothing but the gauge-invariant
{\it physical electron} introduced by Dirac \cite{BookDirac}.
(For more discussion about it, we recommend
the references \cite{Lavelle93A},\cite{Lavelle93B}.) 
Using the same function $\Lambda (x)$,
the gauge-invariant potential $A^\prime_\mu (x)$ can also be readily
found as
\begin{eqnarray}
 \bm{A}^\prime (x) \ &=& \ \bm{A}_\perp (x), \\
 A^{\prime 0} (x) \ &=& \ A^0 (x) \ + \ \int_{- \,\infty}^x \,
 \dot{\bm{A}}_\parallel (x^0, \bm{z}) \cdot d \bm{z} .
\end{eqnarray}
In this way, one reconfirms that the spatial component of the
{\it gauge-invariant potential} $A^\prime_\mu (x)$ is nothing but the
transverse component of $\bm{A} (x)$.

We can show another interesting example in which we can define gauge-invariant
electron and photon fields, which are also path independent at least formally.
The construction begins with introducing a constant 4-vector $n^\mu$.
By using it, we introduce the following decomposition of the photon field : 
\begin{equation}
 A_\mu \ = \ A_\mu^{phys} (x) \ + \ A_\mu^{pure} (x) \ \equiv \ 
 (\,P_{\mu \nu} \ + \ Q_{\mu \nu} \,) \,\,A^\nu (x) \label{Axial_decomp} ,
\end{equation}
where
\begin{eqnarray}
 P_{\mu \nu} \ &=& \ g_{\mu \nu} \ - \ \frac{\partial_\mu \,n_\nu}{n \cdot \partial},
 \label{Axial_proj_phys} \\
 Q_{\mu \nu} \ &=& \ \frac{\partial_\mu \,n_\nu}{n \cdot \partial} . 
 \label{Axial_proj_pure}
\end{eqnarray}
One can verify that the projection operators $P^{\mu \nu}$ and $Q^{\mu \nu}$
satisfies the identities,
\begin{eqnarray}
 P_{\mu \lambda} \,P^\lambda{}_\nu \ &=& \ P_{\mu \nu}, \\
 P_{\mu \lambda} \,Q^\lambda{}_\nu \ &=& \ 
 Q_{\mu \lambda} \,P^\lambda{}_\nu \ = \ 0, \\
 Q_{\mu \lambda} \,Q^\lambda{}_\nu \ &=& \ Q_{\mu \nu} .
\end{eqnarray}
The two components of the above decomposition satisfy the important
properties : 
\begin{eqnarray}
 n^\mu \,A_\mu^{phys} (x) \ &=& \ 0, \label{G_fixing_cond} \\
 \partial_\mu \,A_\nu^{pure} (x) \ - \ \partial_\nu \,A_\mu^{pure} (x) \ &=& \ 0.
 \label{4dim_irrotational}
\end{eqnarray}
As can be easily verified, under a general abelian gauge transformation
$A_\mu (x) \rightarrow A_\mu (x) \,+ \,\partial_\mu \omega (x)$,
these two components respectively transform as
\begin{eqnarray}
 A_\mu^{phys} (x) \ &\rightarrow& \ A_\mu^{phys} (x) , \\
 A_\mu^{pure} (x) \ &\rightarrow& \ A_\mu^{pure} (x) \ + \ 
 \partial_\mu \omega (x) .
\end{eqnarray}
Now we propose to taking
\begin{equation}
 \Lambda (x) \ = \ - \,\int_{- \,\infty}^0 \,A^{pure}_\sigma (z) \,
 \frac{\partial z^\sigma}{\partial \xi} \,d \xi \ = \ 
 - \,\int_{- \,\infty}^x \,A^{pure}_\mu (z) \,d z^\mu , \label{Lambda_axial}
\end{equation}
and define the new electron and photon fields by (\ref{GI_electron})
and (\ref{GI_photon}).
Very interestingly, we can show that the line-integral in the equation above
is actually path-independent.
In fact, let us recall the Stokes' theorem in 4-dimensional space-time
expressed as
\begin{equation}
 \oint \,A_\mu (z) \,d z^\mu \ = \ \frac{1}{2} \,\int_S \,
 \left(\, \partial_\mu \,A_\nu \ - \ \partial_\nu \,A_\mu \,\right)
 d \sigma^{\mu \nu} \, ,
\end{equation}
where $d \sigma^{\mu \nu}$ is an infinitesimal area element tensor.
Owing to the property (\ref{4dim_irrotational}), it holds that
\begin{equation}
 \oint \,A^{pure}_\mu (z) \,d z^\mu \ = \ 0.
\end{equation}
Because of this fact, $\Lambda (x)$ defined by (\ref{Lambda_axial}) is
formally path-independent, and can be expressed as
\begin{eqnarray}
 \Lambda (x) \ &=& \ - \,\int_{- \,\infty}^x \,\,
 \frac{\partial^z_\mu \,n_\nu}{n \cdot \partial^z} \,\,A^\nu (z) \,d z^\mu
 \nonumber \\
 &=& \ - \,\int_{- \,\infty}^x \,\partial^z_\mu \,\left\{\,
 \frac{n \cdot A(z)}{n \cdot \partial^z} \,\right\} \,d z^\mu \ = \ 
 \frac{n \cdot A(x)}{n \cdot \partial} ,
\end{eqnarray}
where $\partial^z_\mu \,\equiv \,\frac{\partial}{\partial z^\mu}$, while
$\partial_\mu \,\equiv \,\frac{\partial}{\partial x^\mu}$.
The gauge-invariant electron and photon fields are therefore given by
\begin{eqnarray}
 \psi^\prime (x) \ &=& \ e^{\,i \,e \,\Lambda (x)} \,\psi (x) \ = \
 e^{\,i \,e \,\frac{n \cdot A(x)}{n \cdot \partial}} \,\psi (x) , \\
 A^\prime_\mu (x) \ &=& \ A_\mu (x) \ + \ \partial_\mu \Lambda (x) \ = \ 
 \left( g_{\mu \mu} \ - \ \frac{\partial_\mu \,n_\nu}{n \cdot \partial} \,
 \,\right) \,A^\nu (x) \ = \ A^{phys}_\mu (x) \label{GI_photon_axial}.
\end{eqnarray}
One notices that the condition $n^\mu \,A^{phys}_\mu = 0$ is nothing but
the gauge-fixing condition projecting out the physical component of the
gauge field in the framework of general axial gauge.

Several remarks are in order here. The familiar gauge fixing condition
$n^\mu \,A_\mu = 0$ does not completely fix the gauge, that is,
there still remains residual gauge degrees of freedom.
The singular nature of the operator $1 \,/\,(n \cdot \partial)$ is
related to these residual degrees of freedom. How to treat this
singularity is connected with what boundary condition is imposed for
the gauge field at the infinity.
Another concern is a generalization to the nonabelian case.
In the abelian case, we have seen that $A^{phys}_\mu (x)$ and $A^{pure}_\mu (x)$
defined by (\ref{Axial_decomp}) supplemented with (\ref{Axial_proj_phys})
and (\ref{Axial_proj_pure}) satisfy the desired gauge
transformation properties.
Unfortunately, the matter is not so simple in the nonabelian gauge case.
In this case, we need more sophisticated method for projecting out
the physical component of the gauge field as discussed in the next section. 

As is clear from the discussion above, except for some fortunate choices
of $\Lambda (x)$, the fields $\psi^\prime (x)$ and $A^\prime_\mu (x)$
defined by (\ref{GI_electron}) and (\ref{GI_photon}) supplemented with
(\ref{gauge_tr_func}) are by construction gauge-invariant but generally
path-dependent. How should we interpret this path-dependence.
Soon after the paper by DeWitt appeared \cite{DeWitt62}, Belinfante conjectured
that a ``path" is just a ``gauge" \cite{Belinfante62}.
He showed that, by averaging over path-dependent potential over the directions
of all straight lines at constant time converging to the point where the
potential is to be calculated, one is led to the potential in the
Coulomb gauge \cite{Belinfante62}.
On the other hand, Rohrlich and Strocchi applied a
similar averaging procedure over covariant path and they obtained
the potential in the Lorentz gauge \cite{RohrlichStrocchi65}.
It was also demonstrated by Yang that, for a simple quantum mechanical system,
the path-dependence is eventually a reflection of
the gauge-dependence \cite{Yang85}. All these investigations
appears to indicate that, if a quantity in question is
seemingly gauge-invariant but path-dependent, it is not a gauge-invariant
quantity in a {\it true} or {\it traditional} sense, which in turn indicates
that it may not correspond to {\it genuine} observables.
Clearly, the GIE approach is equivalent to the standard treatment
of gauge theory, only when its extension by means of gauge link is path-independent.
By the standard treatment of the gauge theory, we mean the following.
Start with a gauge-invariant quantity or expression.
Fix gauge according to the need of practical calculation. 
Answer should be independent of gauge choice.

Now we come back to our original question. We are asking whether the
gluon spin part in the longitudinal nucleon spin sum rule is a gauge-invariant
quantity in a traditional sense or not. In principle, there are two ways to answer
this question. The first is to show that the gauge-invariant longitudinal
gluon spin operator can be constructed without recourse to the notion of ``path".
The second possibility is to adopt a gauge-invariant but generally path-dependent
formulation at the beginning and then to show that the quantity of our
interest is actually path-independent. In the following analysis,
we take the second route and try to show the traditional gauge-invariance
of the evolution equation of the quark and gluon longitudinal spins
in the nucleon.

\section{gauge- and path-independence of the evolution matrix for quark and
gluon longitudinal spins in the nucleon}

A primary question we want to address in this section is whether the gluon
spin term appearing in the longitudinal nucleon spin sum rule is a
gauge-invariant quantity in a traditional sense or whether it is a
quantity that has a meaning only in the light-cone gauge or in the
gauge-invariant extension based on the light-cone gauge.
We have already pointed out that, even for the abelian case, the choice of gauge,
the choice of Lorentz frame,
and the transverse-longitudinal decomposition are all intrinsically interwind. Moreover, an additional complexity arises in the case of nonabelian gauge
theory. The past studies
shows that, different from the abelian gauge theory, even within the
noncovariant treatment, the transverse component cannot be
expressed in a closed form, that is, it can be given only in the form of
perturbation series in the gauge coupling
constant \cite{Lavelle94},\cite{Chen11PLB}.
Still, it remains to be true that the independent dynamical degrees of freedom
of the massless vector field are two transverse components.
In broad terms, one might say that the physics is contained in the
transverse part of the gauge field. In the past, tremendous efforts have been
made to figure out these physical components of the gauge field.
DeWitt's formulation of the electrodynamics explained before is one
typical example \cite{DeWitt62}.
We realized that, especially useful for our purpose is a slightly more
sophisticated formulation proposed in the papers by
Ivanov, Korchemsky and Radyushkin \cite{IK85},\cite{IKR86}.
It is based on the geometric interpretation of the gauge field
actualized as fiber-bundle formulation of gauge theories.
(See \cite{Lorce12B} for a recent concise review on this geometrical formulation.) 
In this approach, the gauge field is identified
with the connection of the principle fiber bundle $M (R^4, G)$
with the 4-dimensional space-time $R^4$ being its base space and
with the fiber being the gauge group $G$.
For the gauge field $A_\mu (x)$ and each element $g (x)$ of the
fiber $G (x)$, one can define the gauge field configuration $A^g (x)$ by
\begin{equation}
 A^g (x) \ = \ g^{-1} (x) \,\left( \,A_\mu (x) \ + \ \frac{i}{g} \,
 \partial_\mu \,\right) \,g (x). \label{Eq:g-trans}
\end{equation}
Then, the set $\left\{ A^g_\mu (x) \right\}$ for all $g (x)$ forms
the gauge equivalent field configurations called the orbits.
For the quantization, one must choose unique gauge orbit from
infinitely many gauge equivalent orbits.
The most popular way of doing it is to impose an appropriate
gauge fixing condition $f ( A^g, g) = 0$ by hand. 
However, the gauge-fixing condition $f (A^g, g) = 0$ sometimes
does not have a unique solution beyond perturbative regime \cite{Gribov78}.
Then, a new method, which is in principle free from the constraints of
perturbative gauge-fixing procedure, was proposed.
In this framework, the gauge function $g (x)$ is fixed as a solution
of the parallel transport equation in the fiber bundle space
\begin{equation}
 \frac{\partial z^\mu}{\partial s} \,D_\mu [A \,] \,g (z(s)) \ = \ 0,
\end{equation}
where $D_\mu = \partial_\mu - i \,g \,A_\mu (z(s))$ is the covariant
derivative, while $z(s)$ is a path $C$ in the 4-dimensional base space
$R^4$ with the following boundary conditions
\begin{equation}
 z^\mu (s = 1) \ = \ x^\mu, \ \ \ z^\mu (s = 0) \ = \ x^\mu_0.
\end{equation}
The solution to this equation is well known. It is expressed in terms of
the Wilson line as
\begin{equation}
 g (x) \ = \ W_C (x, x_0) \,g (x_0) ,
\end{equation}
with
\begin{equation}
 W_C (x, x_0) \ = \ P \,\exp \,\left[\, i \,g \,\int_{x_0}^x \,
 d z^\mu \,A_\mu (z) \,\right] . \label{Eq:Wilson-line}
\end{equation}
Once $g (x)$ is given, $A^g_\mu (x)$ defined by (\ref{Eq:g-trans}) is
uniquely specified. However, one
should clearly keep in mind the fact that $A^g_\mu (x)$ so
determined is generally dependent on the choice of path $C$ connecting
$x$ and $x^0$ (the starting point of the path).
By substituting (\ref{Eq:Wilson-line}) into (\ref{Eq:g-trans}) and
by using the derivative formula for the Wilson line, together with
the identity $W^{-1}_C (x,y) = W_C (y,x)$, $A^g_\mu (x)$ can be expressed as
\begin{eqnarray}
 A^g_\mu (x) \ &=& \ \ A_\nu (x_0) \,\,
 \frac{\partial x^\nu_0}{\partial x^\mu}
 \nonumber \\
 \ &-& \ 
 \int_{x_0}^x \,d z^\nu \,\,
 \frac{\partial z^\rho}{\partial x^\mu} \,\,W_C (x_0, z) \,
 F_{\rho \nu} (z \,; A) \,W_C (z,x_0) , \label{Eq:projection}
\end{eqnarray}
where $F_{\nu \rho} (z \,;\,A) \equiv \partial_\nu \,A_\rho (z) - 
\partial_\rho \,A_\nu (z) - \,i \,g \,[A_\nu (z), A_\rho (z)]$.
The r.h.s. of the above equation depends on the
original gauge field $A_\mu$, and on the starting point $x_0$ of the path,
which in principle can depend on $x$.
In the following, we take $x_0$ to be a unique point for all contours $C$,
so that $\partial x^\nu_0 \,/\,\partial x^\mu = 0$.

With some natural constraints on the choice of the contours
$C$, it was shown in \cite{IK85},\cite{IKR86} (see also \cite{SS98}) that
the above way of fixing the gauge is equivalent to taking gauges
satisfying the condition
\begin{equation}
 W_C (x, x_0) \ = \ P \,\exp \,\left[\, i \,g \,
 \int_{x_0}^x \,d z^\mu \,A^g_\mu (z) \,\right] \ = \ 1 .
\end{equation}
This class of gauges are called the {\it contour gauges}. 
An attractive feature of the contour gauge is that they are ghost-free.
As specific examples of contour gauges, they briefly discussed three
gauges.
They are the Fock-Schwinger gauge, the Hamilton gauge, and the general
axial gauge. Especially useful for our purpose here is the general axial gauge.
The reason is that this is the most convenient gauge among the three
for perturbative calculations. In the context of geometrical approach,
the axial gauge corresponds to taking the following infinitely long
straight-line path,
\begin{equation}
 z^\mu (s) \ = \ x^\mu \ + \ s \,n^\mu ,
\end{equation}
with $0 < s < \infty$, where $n^\mu$ is a constant 4-vector characterizing
the direction of the path.
Substituting this form of $z^\mu (s)$ into (\ref{Eq:projection}),
one obtains the relation
between the transformed and original gauge fields as
\begin{equation}
 A^g_\mu (x) \ = \ n^\nu \,\int_0^\infty \,d s \,\,
 W_C^\dagger (x + n \,s, \infty) \,
 F_{\mu \nu} (x + n \,s \,;\,A) \,W_C (x + n \,s, \infty) ,
 \label{Eq:axial-projection}
\end{equation}
with
\begin{equation}
 W_C (x, \infty) \ = \ P \,\exp \,\left(\,i \,g \,
 \int_0^\infty \,d s \,\,n^\mu \,A_\mu (x + n \,s) \,\right) .
\end{equation}
Taking account of the antisymmetry of the field strength tensor, 
$F_{\nu \mu} = - \,F_{\mu \nu}$, one can easily convince that $A^g_\mu$
satisfies the identity :
\begin{equation}
 n^\mu \,A^g_\mu \ = \ 0 .
\end{equation}
Note that this is nothing but the gauge fixing condition in the general axial gauge.
Since $n^\mu$ is an arbitrary constant 4-vector, this class of gauge
contains several popular gauges. For instance, by choosing as
$n^\mu = (1,0,0,0)$, $n^\mu = (1,0,0,1) \,/\,\sqrt{2}$, and $n^\mu = (0,0,0,1)$,
we can cover any of the temporal gauge, the light-cone gauge,
and the spatial axial gauge.
Furthermore, using the property of the Wilson line
\begin{equation}
 W_C^\dagger (x + n \,s, \infty) \,F_{\mu \nu} (x + n \,s \,;\, A) \,
 W_C (x + n \,s, \infty)
 \ = \ F_{\mu \nu} (x + n \,s \,;\, A^g) ,
\end{equation}
Eq.(\ref{Eq:axial-projection}) can also be expressed in an equivalent but
simpler form as
\begin{equation}
 A^g_\mu (x) \ = \ n^\nu \,\int_0^\infty \,d s \,\,F_{\mu \nu} (x + n \,s \,;\,A^g) .
\end{equation}
This identity represents the fact that, in the general axial gauge,
the gauge potential $A_\mu$ can be expressed in terms of the field-strength
tensor \cite{BB91},\cite{ZhangPak12}.
(Undoubtedly, Ivanov et al. correctly recognized the fact that the choice of a
path in the geometrical formulation just corresponds to a
gauge-fixing procedure. Note that this understanding is nothing different
from the conclusion of Belinfante pointed out before that a ``path'' is just a
``gauge''.)
With the identification $A^{phys}_\mu (x) \equiv A^g_\mu (x)$, the above equation
can then be thought of as a defining equation of the physical component
$A^{phys}_\mu (x)$ of the gluon field based on the general axial gauge.
We emphasize that this defining
equation itself is free from perturbation theory in the gauge coupling constant.

Since the main purpose of our present study is to show the perturbative
gauge-invariance of the gluon spin, or more concretely, the traditional
gauge-invariance of the evolution equation of the quark and gluon
longitudinal spins in the nucleon, let us look into the perturbative
contents of the above equality (\ref{Eq:axial-projection}),
which can be interpreted as an equation projecting out the physical
component of the gluon field $A^{phys}_\mu \equiv A^g_\mu (x)$
from $A_\mu (x)$.
At the lowest order in the gauge coupling constant, this gives
\begin{equation}
 A^{phys}_\mu (x) \ \simeq \ n^\nu \,
 \int_0^\infty \,d s \,\left(\,\partial_\mu \,A_\nu (x + n \,s) \ - \ 
 \partial_\nu \,A_\mu (x + n \,s) \,\right) . 
\end{equation}
Introducing the Fourier transform of $A_\mu (x)$,
\begin{equation}
 \tilde{A}_\mu (k) \ = \ \int \,d^4 x \,e^{\,- \,i \,k \,x} \,
 A_\mu (x) ,
\end{equation}
we therefore get
\begin{eqnarray}
 A^{phys}_\mu (x) \ &\simeq& \ n^\nu \,\int_0^\infty \,d s \,
 \int \,\frac{d^4 k}{(2 \,\pi)^4} \,\,e^{\,i \,k \,(x + n \,s)} \,
 \left(\, i \,k_\mu \,\tilde{A}_\nu (k) \ - \ 
 i \,k_\nu \,\tilde{A}_\mu (k) \,\right) \nonumber \\
 &=& \ \int \,\frac{d^4 k}{(2 \,\pi)^4} \,\,e^{\,i \,k \,x} \,
 \left(\, g_{\mu \nu} \ - \ \frac{k_\mu \,n_\nu}{k \cdot n} \,\right) \,
 \tilde{A}^\nu (k) \nonumber \\
 \ &=& \ \left(\, g_{\mu \nu} \ - \ 
 \frac{\partial_\mu \,n_\nu}{n \cdot \partial} \,\right) \,A^\nu (x).
\end{eqnarray}
Note that, although this is simply the lowest order expression for the
physical component for $A_\mu (x)$ in the case of nonabelian gauge theory,
it reproduces the exact one (\ref{GI_photon_axial}) in the abelian case,
discussed in the previous section. 
One can easily verify that this
gives the lowest order expression for the physical propagator of the gluon
as
\begin{equation}
 \langle \,T \,(A^{phys}_{\mu, a} (x) \,A^{phys}_{\nu, b} (x)) \,\rangle^{(0)}
 \ = \ 
 \int \,\frac{d^4 k}{(2 \,\pi)^4} \,\,e^{\,i \,k \,(x - y)} \,\,
 \frac{- \,i \,\delta_{a b}}{k^2 + i \,\epsilon} \,\,P_{\mu \nu} (k)
\end{equation}
with
\begin{equation}
 P_{\mu \nu} (k) \ = \ g_{\mu \nu} \ - \ 
 \frac{k_\mu \,n_\nu \,+ \,n_\mu \,k_\nu}{k \cdot n} \ + \ 
 \frac{n^2 \,k_\mu \,k_\nu}{(k \cdot n)^2} .
\end{equation}
As anticipated, it just coincides with the free gluon propagator in the
general axial gauge.

In this way, one finds that the path dependence or direction dependence of
the constant 4-vector $n^\mu$ in the geometrical formulation is replaced by the
gauge dependence within the class of gauges called the general axial gauge.
In this setting, then, the gluon spin operator reduces to
$M^{\lambda \mu \nu}_{G-spin} \, = \, 2 \,\mbox{Tr} \,[\,
F^{\lambda \nu} \,A^\mu \, - \, F^{\lambda \mu} \,A^\nu \,]$,
where $A_\mu$ in this equation should be regarded as the gluon field satisfying
the general axial gauge condition $n^\mu \,A_\mu = 0$.

Our strategy should be clear by now. We want to investigate the 1-loop
anomalous dimension for the quark and gluon longitudinal spin operators in the
nucleon within the general axial gauge characterized by the 4-vector $n^\mu$.
Since the general axial gauge falls into a category of the so-called
noncovariant gauges, one must be careful about the
fact that the choice of gauge and the choice of Lorentz-frame are
intrinsically interwined. To understand this subtlety, it is instructive to
remember the basis of the longitudinal momentum sum rule of the nucleon.
The momentum sum rule of the nucleon is derived based on the following
covariant relation,
\begin{equation}
 \langle P s \,|\, T_{\mu \nu} (0) \,|\, P s \rangle \ = \ 
 2 \,P_\mu \,P_\nu , \label{Eq:Emom_covariant}
\end{equation}
where $T_{\mu \nu}$ is the (symmetric) QCD energy momentum tensor,
while $|P s \rangle$ is a nucleon state with momentum $P$ and spin $s$.
A useful technique for obtaining the momentum sum rule is to
introduce a light-like constant vector $n^\mu$ with $n^2 = 0$.
By contracting (\ref{Eq:Emom_covariant}) with $n^\mu$ and $n^\nu$, we have
\begin{equation}
 \frac{\langle P s \,|\, n^\mu \,T_{\mu \nu} (0) \,n^\nu \,|\, P s \rangle}
 {2 \,(P^+)^2} \ = \ 1,
\end{equation}
which provides us with a convenient basis for obtaining a concrete form
of the momentum sum rule of QCD.
Since (\ref{Eq:Emom_covariant}) itself is relativistically
covariant, the above choice of $n^\mu$ is not an only choice, however.
With the choice of arbitrary constant four-vector $n^\mu$
with $n^2 \neq 0$, we would have more general relation,
\begin{equation}
 \frac{\langle P s \,|\, n^\mu \,T_{\mu \nu} (0) \,n^\nu \ - \ 
 \frac{1}{4} \,n^2 \,T^\mu{}_\mu (0) \,|\, P s \rangle}
 {2 \,(P \cdot n)^2} \ = \ 1, \label{Eq:Emom_noncovariant}
\end{equation}
Here, since $n^2 \neq 0$, the subtraction of the trace term is obligatory.

Similarly, the starting point for obtaining the longitudinal nucleon
spin sum rule is the following covariant relation : 
\begin{equation}
 \langle P s \,|\, M^{\lambda \mu \nu} (0) \,|\, P s \rangle \ = \ 
 J_N \,\,\frac{P_\rho \,s_\sigma}{M_N^2} \,\,
 [\, 2 \,P^\lambda \,\,\epsilon^{\nu \mu \rho \sigma} \ - \ 
 P^\mu \,\,\epsilon^{\lambda \nu \rho \sigma} \ - \ 
 P^\nu \,\,\epsilon^{\mu \lambda \rho \sigma} \,] ,
\end{equation}
where $M^{\lambda \mu \nu}$ is the angular momentum tensor of QCD, while 
\begin{equation}
 P^2 \ = \ M_N^2, \ \ \ s^2 \ = \ - \,M_N^2, \ \ \ s \cdot P \ = \ 0,
\end{equation}
with $s_\mu$ being a covariant spin-vector of the nucleon.
Note that, without loss of generality, we can take as
$P^\mu \ = \ (P^0, 0, 0, P^3)$ and $s^\mu \ = \ (P^3, 0, 0, P^0)$
with $P^0 = \sqrt{(P^3)^2 + M_N^2}$.
The longitudinal nucleon spin sum rule can be obtained by
setting $\mu = 1, \nu = 2$, which gives
\begin{eqnarray}
 \langle P s \,|\, M^{\lambda 1 2} (0) \,|\, P s \rangle \ = \ 
 - \,2 \,J_N \,\frac{1}{M_N^2} \,\,P^\lambda \,\epsilon^{1 2 \rho \sigma} \,
 P_\rho \,s_\sigma \ = \ 
 2 \,J_N \,P^\lambda . \label{Eq:Spin_tensor}
\end{eqnarray}
Contracting this relation with an arbitrary constant 4-vector $n_\lambda$,
we therefore arrive at the basis equation of the longitudinal nucleon spin
sum rule \cite{JM90} : 
\begin{equation}
 J_N \ = \ \frac{1}{2} \ = \ 
 \frac{\langle P s \,|\, n_\lambda \,M^{\lambda 1 2}(0) \,|\,P s \rangle}
 {2 \,(P \cdot n)} . \label{Eq:Spin_noncovariant}
\end{equation}
An important fact here is that the relations (\ref{Eq:Emom_noncovariant})
and (\ref{Eq:Spin_noncovariant}) are not covariant ones
any more. The 4-vector $n^\nu$ appearing in these equations should therefore
be identified with the 4-vector that characterizes the Lorentz-frame, in which
the gauge-fixing condition $n^\mu \,A_\mu = 0$ is
imposed \cite{BookBD65}.

In the following, we shall confine to the intrinsic spin parts of quarks
and gluons appearing in the nucleon spin decompositions (we recall the fact
that they are just common in both decompositions (I) and (II)) :
\begin{eqnarray}
 M^{\lambda \mu \nu}_{q-spin} \ &=& \ \frac{1}{2} \,\,
 \epsilon^{\lambda \mu \nu \sigma} \,\,
 \bar{\psi} \,\gamma_\sigma \,\gamma_5 \,\psi, \\
 M^{\lambda \mu \nu}_{G-spin} \ &=& \ 2 \,\,\mbox{Tr} \,
 [\,F^{\lambda \nu} \,A^\mu
 \ - \ F^{\lambda \mu} \,A^\nu \,] .
\end{eqnarray}
Here, the gauge fields appearing in $M^{\lambda \mu \nu}_{G-spin}$ should
be regarded as the physical gluon field satisfying the general axial-gauge
condition $n^\mu \,A_\mu = 0$.

Generally, the gluon spin operator appearing in (\ref{Eq:Spin_tensor})
consists of three pieces as
\begin{eqnarray}
 M^{\lambda 1 2}_{G-spin} \ &=& \ 2 \,\,\mbox{Tr} \,[\,F^{\lambda 1} \,A^2
 \ - \ F^{\lambda 2} \,A^1 \,] \ = \ 
 V_A^\lambda \ + \ V_B^\lambda \ + \ V_C^\lambda , 
 \label{Eq:Gluon_spin}
\end{eqnarray}
where
\begin{eqnarray}
 V_A^\lambda \ &=& \ (\partial^\lambda \,A^1_a) \,A^2_a \ - \ 
 (\partial^\lambda \,A^2_a) \,A^1_a , \label{Eq:Gluon_spin_A} \\
 V_B^\lambda \ &=& \ - \,[\,(\partial^1 \,A^\lambda_a \,A^2_a \ - \ 
 (\partial^2 \,A^\lambda_a) \,A^1_a \,] , \label{Eq:Gluon_spin_B} \\
 V_C^\lambda \ &=& \ g \,f_{a b c} \,A^\lambda_b \,\,[\,A^1_c \,A^2_a \ - \ 
 A^2_c \,A^1_a \,] . \label{Eq:Gluon_spin_C}
\end{eqnarray}
Note however that $V_B^\lambda$ and $V_C^\lambda$ terms do not
contribute to the longitudinal nucleon spin sum rule (\ref{Eq:Spin_noncovariant}),
since $n_\lambda \,V_B^\lambda = n_\lambda \,V_C^\lambda = 0$, due to the
gauge-fixing condition $n_\lambda \,A^\lambda = 0$.
As a consequence, in the general axial gauge, only the $V_A^\lambda$ term
contributes to the longitudinal spin sum rule.

\vspace{5mm}
\begin{figure}[ht]
\begin{center}
\includegraphics[width=6.5cm]{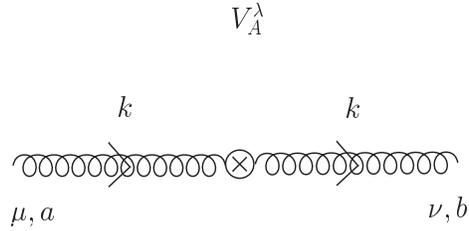}
\caption{Momentum space vertices for the gluon spin.}
\label{Fig1_Gspin_Vertex}
\end{center} 
\end{figure}

The momentum space vertex for the gluon spin therefore reduces to the
following simple form supplemented with the diagram illustrated
in Fig.\ref{Fig1_Gspin_Vertex} :
\begin{eqnarray}
 V_A^\lambda \ &=& \ 2 \,i \,k^\lambda \,(\,g^{\mu 1} \,g^{\nu 2} \,- \,
 g^{\mu 2} \,g^{\nu 1} \,) \,\delta_{a b}.
\end{eqnarray}

Now we are ready to investigate the anomalous dimension matrix for the
longitudinal quark and gluon spins in the nucleon,
\begin{equation}
 \Delta \gamma \ = \ \left( \begin{array}{cc}
 \Delta \gamma_{qq} & \Delta \gamma_{qG} \\
 \Delta \gamma_{Gq} & \Delta \gamma_{GG} \\
 \end{array}
 \right) ,
\end{equation}
which controls the scale evolution of the quark and gluon spins.
We start with the quark spin operator $M^{\lambda 1 2}_{q-spin}$, although there
is no known problem in this part. The reason is that we want to convince the
independence of the final result on the choice of the constant 4-vector $n^\mu$,
which specifies the Lorentz frame in which the gauge-fixing
condition necessary for the quantization of the gluon field is imposed.

\vspace{5mm}
\begin{figure}[ht]
\begin{center}
\includegraphics[width=6.0cm]{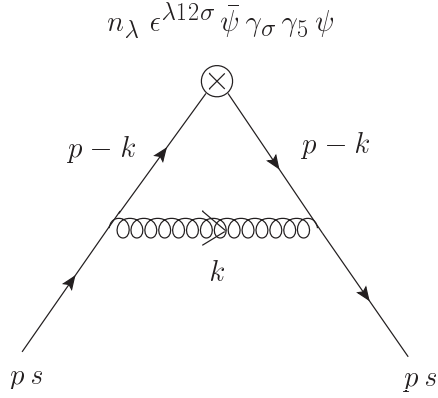}
\caption{The Feynman diagram contributing to $\Delta \gamma_{qq}$.}
\label{Fig2_Vertex_qq}
\end{center} 
\end{figure}

The anomalous dimension $\Delta \gamma_{qq}$ can be obtained by evaluating
the matrix element of
\begin{equation}
 2 \,n_\lambda \,M^{\lambda 1 2}_{q-spin} \ = \ 
 n_\lambda \,\epsilon^{\lambda 1 2 \sigma} \,
 \bar{\psi} \,\gamma_\sigma \,\gamma_5 \,\psi ,
\end{equation}
in a longitudinally polarized quark state $|\,p s \rangle$ with $s = \pm \, 1$.
The corresponding 1-loop diagram is shown in Fig.\ref{Fig2_Vertex_qq}.
This gives
\begin{eqnarray}
 T_{qq} \ &=& \ \frac{1}{2 \,p \cdot n} \,\int \,\frac{d^4 k}{(2 \,\pi)^4} \,\,
 \bar{u} (ps) \,i \,g \,\gamma^\nu \,t^a \,
 \frac{i \,(\displaystyle{\not} p - \displaystyle{\not} k)}
 {(p - k)^2 + i \,\varepsilon} \,\,n_\lambda \,\,
 \epsilon^{\lambda 1 2 \sigma} \,\gamma_\sigma \,\gamma_5 \nonumber \\
 &\,& \hspace{20mm} \times \ 
 \frac{i \,(\displaystyle{\not} p - \displaystyle{\not} k)}
 {(p-k)^2 + i \,\varepsilon} \,\,i \,g \,\gamma^\mu \,t^b \,\,u (ps) \,\,
 \delta_{a b} \,D_{\mu \nu} (k) , 
\end{eqnarray}
where
\begin{equation}
 D_{\mu \nu} (k) \ = \ \frac{- \,i}{k^2 + i \,\varepsilon} \,\,P_{\mu \nu} (k)
\end{equation}
with
\begin{equation}
 P_{\mu \nu} (k) \ \equiv \ P_{\mu \nu}^{axial} (k) \ = \ 
 g_{\mu \nu} \ - \ \frac{k_\mu \,n_\nu + n_\mu \,k_\nu}{k \cdot n} \ + \ 
 \frac{n^2 \,k_\mu \,k_\nu}{(k \cdot n)^2} .
\end{equation}
As is well-known, this gluon propagator in the general axial gauge contains
a spurious simple pole and also a double pole. 
In the following, let us evaluate the contributions of the three terms in
$P^{axial}_{\mu \nu} (k)$ separately. The calculation of the part containing
$g_{\mu \nu}$ is straightforward. After some Dirac algebra, we get
\begin{eqnarray}
 T_{qq} (g_{\mu \nu}) \ &=& \ - \,i \,\,\frac{g^2 \,C_F}{2 \,p \cdot n}
 \ \times \ \left\{\,- \,8 \,n_\alpha \,p_\beta \,
 \int \,\frac{d^4 k}{(2 \,\pi)^4} \,\,\frac{k^\alpha \,k^\beta}
 {[(p - k)^2 + i \,\varepsilon]^2 \,(k^2 + i \,\varepsilon)} \right. 
 \ \ \ \ \nonumber \\
 &\,& \left. \hspace{30mm} + \ 4 \ p \cdot n \,
 \int \,\frac{d^4 k}{(2 \,\pi)^4} \,\,\frac{1}{[(p-k)^2 + i \,\varepsilon]^2} \,
 \right\} .
\end{eqnarray}
Using the standard dimensional regularization with $D \equiv 2 \,\omega$
space-time dimension, the divergent parts of the necessary integral are given by
\begin{eqnarray}
 &\,& \mbox{div} \,\int \,\frac{d^4 k}{(2 \,\pi)^4} \,\frac{k^\alpha \,k^\beta}
 {[(p-k)^2 + i \,\varepsilon]^2 \,(k^2 + i \,\varepsilon)} \ = \ 
 \frac{1}{4} \,\,g^{\alpha \beta} \,\,\bar{I} , \\
 &\,& \mbox{div} \,\int \,\frac{d^4 k}{(2 \,\pi)^4} \,\frac{1}{[(p-k)^2 + i \,\varepsilon]^2}
 \ = \ \bar{I} ,
\end{eqnarray}
where
\begin{equation}
 \bar{I} \ = \ \frac{i \,\pi^2}{2 - \omega} .
\end{equation}
We therefore obtain
\begin{equation}
 T_{qq} (g_{\mu \nu}) \ = \ \frac{\alpha_S}{2 \,\pi} \,\,\,\frac{1}{2} \,\,C_F \,\,
 \frac{1}{2 - \omega} .
\end{equation}

Next, we evaluate the term containing a simple spurious pole
$1 \,/\,(k \cdot n)$. After some algebra, we get
\begin{eqnarray}
 T_{qq} (1 \,/\,(k \cdot n)) \ &=& \ 
 - \,\,i \,\,\frac{g^2 \,C_F}{2 \,p \cdot n} \,\,
 \left\{\, - \,p \cdot n \,\int \,\frac{d^4 k}{(2 \,\pi)^4} \,\,
 \frac{1}{[(p-k)^2 + i \,\varepsilon]^2} \right. \nonumber \\
 &\,& \hspace{24mm} \left. +  \ n^2 \,p_\beta \,\int \,\frac{d^4 k}{(2 \,\pi)^4} \,\,
 \frac{k^\beta}{[(p-k)^2 + i \,\varepsilon]^2 \,k \cdot n} \,\right\} .
\end{eqnarray}
Now we encounter a Feynman integral containing a spurious pole.
A consistent method for handling such Feynman integrals was first proposed
by Mandelstam \cite{Mandelstam83} and independently
by Leibbrandt \cite{Leibbrandt84} in the light-cone gauge corresponding
to the choice $n^2 = 0$.
It is given as
\begin{equation}
 \frac{1}{k \cdot n} \ \rightarrow \ \frac{1}{[k \cdot n]} \ \equiv \ 
 \lim_{\varepsilon \rightarrow 0} \,
 \frac{k \cdot n^*}{k \cdot n \,k \cdot n^* + i \,\varepsilon}, \ \ \ \ \ 
 (\varepsilon > 0)
\end{equation}
where $n^*_\mu = (n_0, - \,\bm{n})$ is a dual 4-vector to the 4-vector
$n_\mu = (n_0, \bm{n})$ with $n^2 = 0$ and $n^{* 2} = 0$.
(Practically, we can take as $n^\mu \ = \ (n^0, 0, 0, n^3)$ and 
$n^{* \mu} \ = \ (n^0, 0, 0, - \,n^3)$ without loss of generality.)
Later, Gaigg et. al. showed that this prescription can be generalized to more
general case of $n^2 \neq 0$ and $n^{* 2} \neq 0$ \cite{Gaigg88A},\cite{Gaigg88B}. 
(For review, see \cite{BassettoBook},\cite{LeibbrandtBook}.)
In this generalized $n^*_\mu$-prescription, the divergent part of the above
integral is given by
\begin{equation}
 \int \,d^{2 \,\omega} k \,\,\frac{k^\beta}
 {[(p-k)^2 + i \,\varepsilon]^2 \,[k \cdot n]} \ = \ 
 \frac{1}{D} \,\left(\,n^{* \beta} \ - \ \frac{n^{*2}}{n^* \cdot n + D} \,\,n^\beta
 \,\right) \,\bar{I} ,
\end{equation}
where $D$ is defined by
\begin{equation}
 D \ \equiv \ \sqrt{(n^* \cdot n)^2 \ - \ n^{* 2} \,n^2} .
\end{equation}
By using this result, the divergent part of $T_{qq} (1 \,/\,(k \cdot n))$ becomes
\begin{eqnarray}
 T_{qq} (1 \,/\,(k \cdot n)) &=& \frac{\alpha_S}{4 \,\pi} \,\,C_F \,\,
 \frac{1}{2 - \omega} \nonumber \\
 &\,& \times \ \left\{\,- \,\,4 \ + \ 2 \,\,\frac{n^2}{[p \cdot n]} \,\,\frac{1}{D} \,
 \left( p \cdot n^* \ - \ \frac{n^{* 2}}{n^* \cdot n + D} \,\,p \cdot n \,\right)  
 \,\right\} .
\end{eqnarray}

The contribution of the part containing a spurious double pole structure
$1 \,/\, (k \cdot n)^2$ can similarly be calculated. We get
\begin{equation}
 T_{qq} (1 \,/\,(k \cdot n)^2) \ = \ - \,i \,g^2 \,C_F \,n^2 \,
 \int \,\frac{d^4 k}{(2 \,\pi)^4} \,\,
 \frac{k^2}{[(p-k)^2 + i \,\varepsilon]^2 \,(k \cdot n)^2} .
\end{equation}
Using the generalized $n^*_\mu$-prescription again, the divergent part of the
relevant integral is given by
\begin{equation}
 \mbox{div} \,\int \,\frac{d^4 k}{(2 \,\pi)^4} \,
 \frac{k^2}{[(p-k)^2 + i \,\varepsilon]^2 \,[k \cdot n]^2} \ = \ 
 \frac{2}{D} \,\,\frac{n^{* 2}}{n^* \cdot n + D} \,\,\bar{I} .
\end{equation}
We therefore obtain
\begin{equation}
 T_{qq} \,(1 \,/\,(k \cdot n)^2) \ = \ - \,\,2 \,\,\frac{\alpha_S}{4 \,\pi} \,\,C_F \,
 \left(\,1 \ - \ \frac{n \cdot n^*}{D} \,\right) \,\frac{1}{2 - \omega} .
\end{equation}
Here, use has been made of the identity,
\begin{equation}
 \frac{n^2 \,n^{* 2}}{n^* \cdot n + D} \ = \ 
 n \cdot n^* \ - \ D .
\end{equation}
Summing up the three terms, we arrive at
\begin{eqnarray}
 T_{qq} \ &=& \ 
 - \,\frac{\alpha_S}{4 \,\pi} \,\,C_F \,\,\frac{1}{2- \omega} \nonumber \\
 &\,& \,- \ \frac{\alpha_S}{4 \,\pi} \,\,2 \,\,C_F \,\,
 \frac{1}{[p \cdot n]} \,\,\frac{1}{D} \,
 \,[\,p \cdot n \,n \cdot n^* \ - \ p \cdot n^* \,n^2 \,] \,\,\frac{1}{2 - \omega}
 \nonumber \\
 &\,& \,- \ \frac{\alpha_S}{4 \,\pi} \,\,2 \,\,C_F \,\,
 \left( 1 - \frac{n \cdot n^*}{D} \,\right) \,
 \frac{1}{2 - \omega} .
\end{eqnarray}

At this stage, it is instructive to consider several special choices of $n^\mu$.
The light-cone gauge choice corresponds to taking $n^0 = n^3 = 1 \,/\,\sqrt{2}$.
In this case, we have
\begin{equation}
 p \cdot n \ = \ p^+, \ \ p \cdot n^* \ = \ p^-, \ \  n \cdot n^* \ = \ 1,
 \ \ n^2 \ = \ 0,
\end{equation}
and
\begin{equation}
 D \ = \ 1 ,
\end{equation}
so that we find that
\begin{equation}
 T_{qq} (LC) \ = \ - \,\frac{\alpha_S}{2 \,\pi} \,\,\frac{3}{2} \,\,C_F \,\,
 \frac{1}{2 - \omega} .
\end{equation}
This legitimately reproduces the answer first obtained by Ji, Tang and Hoodbhoy
in the light-cone gauge \cite{JTH96}.

Another interesting choice is the temporal gauge limit specified by
$n^0 = 1$ and $n^3 = 0$. In this limit, we have
\begin{equation}
 p \cdot n \ = \ p \cdot n^* \ = \ p^0, \ \ n \cdot n^* \ = \ 1, \ \ 
 n^2 \ = \ 1
\end{equation}
and
\begin{equation}
 D \ = \ 0.
\end{equation}
We therefore find that the coefficients of $1 \,/\,(2 - \omega)$ in the 2nd and
3rd term of $T_{qq}$ {\it diverge}.
The temporal gauge limit is {\it singular} in this respect.
However, for obtaining the anomalous dimension $\Delta \gamma_{qq}$, we must also
take account of the self-energy insertion to the external quark lines.
The contribution of these diagrams can easily be obtained by
using the known result for the 1-loop quark self-energy in the general axial
gauge. (See, for instance, \cite{BassettoBook}). We get
\begin{eqnarray}
 T_{qq}^{Self} \ &=& \ \ \ 
 \frac{\alpha_S}{4 \,\pi} \,\,C_F \,\,\frac{1}{2 - \omega} 
 \nonumber \\
 &\,& + \ \frac{\alpha_S}{4 \,\pi} \,\,2 \,\,C_F \,\,
 \frac{1}{[p \cdot n]} \,\,\frac{1}{D} \,
 [\,p \cdot n \,n \cdot n^* \ - \ p \cdot n^* \,n^2 \,] \,\frac{1}{2 - \omega}
 \nonumber \\
 &\,& + \ \frac{\alpha_S}{4 \,\pi} \,\,2 \,\,C_F \,\,
 \left( 1 - \frac{n \cdot n^*}{D} \,\right) \,
 \frac{1}{2 - \omega} .
\end{eqnarray}
As anticipated, this exactly cancels $T_{qq}$ obtained above, thereby being led
to the standardly-known answer, i.e.
\begin{equation}
 \Delta \gamma_{qq} \ = \ 0 . 
\end{equation}
It is important to recognize that this final result is obtained totally
independently of the choice of the 4-vector $n^\mu$.

\vspace{5mm}
\begin{figure}[ht]
\begin{center}
\includegraphics[width=6.5cm]{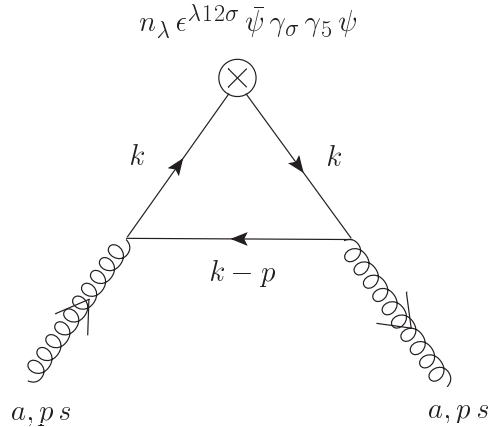}
\caption{The Feynman diagram contributing to $\Delta \gamma_{qG}$.}
\label{Fig3_Vertex_qG}
\end{center} 
\end{figure}

The relevant Feynman diagram contributing to the anomalous dimension
$\Delta \gamma_{q G}$ is illustrated in Fig.\ref{Fig3_Vertex_qG}.
Since no internal gluon propagator appears in this diagram, we do not need
to repeat the standard manipulation.
One can easily verify that
\begin{equation}
 \Delta \gamma_{q G} \ = \ 0.
\end{equation}

\vspace{5mm}
\begin{figure}[ht]
\begin{center}
\includegraphics[width=6.5cm]{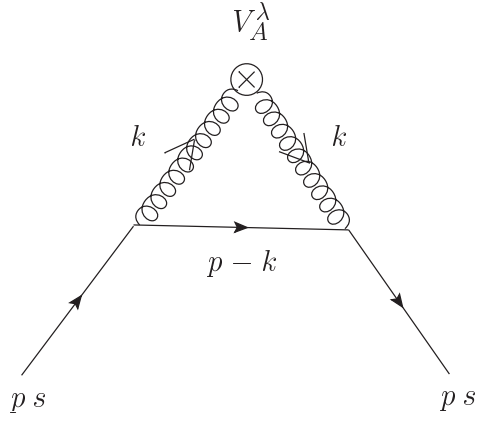}
\caption{The Feynman diagram contributing to $\Delta \gamma_{Gq}$.}
\label{Fig4_Vertex_Gq}
\end{center} 
\end{figure}

Next, we turn to the anomalous dimension $\Delta \gamma_{Gq}$.
The relevant 1-loop Feynman diagram is shown in Fig.\ref{Fig4_Vertex_Gq}.
The contribution of the vertex $V_A$ in the gluon spin operator is given by
\begin{eqnarray}
 T^A_{Gq} \ &=& \ \frac{1}{2 \,p \cdot n} \,\int \,\frac{d^4 k}{(2 \,\pi)^4} \,\,
 \bar{u} (p s) \,i \,g \,\gamma^{\nu^\prime} \,t^d \,\,
 \frac{i \,(\displaystyle{\not} p - \displaystyle{\not} k )}{(p-k)^2 + i \,\varepsilon}
 \,\,\delta^{d e} \nonumber \\
 &\,& \times \ 2 \,i \,(k \cdot n) \,[\, g^{\mu 1} \,g^{\nu 2}
 \ - \ g^{\mu 2} \,g^{\nu 1} \,] \,\,\delta^{b c} \,\,
 i \,g \,\gamma^{\mu^\prime} \,t^e \,\,u (p s) \nonumber \\
 &\,& \times \ \delta^{b d} \,D_{\mu \mu^\prime} (k) \,\,
 \delta^{c e} \,D_{\nu \nu^\prime} (k),
\end{eqnarray}
where $D_{\mu \mu^\prime} (k)$ and $D_{\nu \nu^\prime} (k)$ are gluon
propagators in the general axial gauge excluding trivial color dependent parts.
This gives
\begin{eqnarray}
 T^A_{Gq} \ &=& \ 
 - \,\,\frac{g^2 \,C_F}{p \cdot n} \,\int \,\frac{d^4 k}{(2 \,\pi)^4} \,\,
 \frac{k \cdot n}{(k^2 + i \,\varepsilon)^2 \,[(k-p)^2 + i \,\varepsilon]} \nonumber \\
 &\,& \times \ \bar{u} (p s) \,\gamma^{\nu^\prime} \,
 (\displaystyle{\not} p - \displaystyle{\not} k) \,\gamma^{\mu^\prime} \,u (p s) \,\,
 (\,g^{\mu 1} \,g^{\nu 2} \ - \ g^{\mu 2} \,g^{\nu 1} \,) \,\,
 P_{\mu \mu^\prime} (k) \,\,P_{\nu \nu^\prime} (k) \nonumber \\
 \ &=& \ \,2 \,i \,\,\,\frac{g^2 \,C_F}{p \cdot n} \,\int \,
 \frac{d^4 k}{(2 \,\pi)^4} \,\,
 \frac{k \cdot n}{(k^2 + i \,\varepsilon)^2 \,[(k-p)^2 + i \,\varepsilon]} \nonumber \\
 &\,& \times \ (\,g^{\mu 1} \,g^{\nu 2} \ - \ g^{\mu 2} \,g^{\nu 1} \,) \,\,
 \epsilon^{\mu^\prime \,\nu^\prime \alpha \,\beta} \,\,k_\alpha \,\,p_\beta \,\,
 P_{\mu \mu^\prime} (k) \,\,P_{\nu \nu^\prime} (k) .
\end{eqnarray}
After some algebra, we obtain
\begin{eqnarray}
 T^A_{Gq} \ &=& \ - \,\,4 \,\,i \,\,\frac{g^2 \,C_F}{p \cdot n} \,\,
 \left\{\, \,p_\mu \,n_\nu \,\int \,\frac{d^4 k}{(2 \,\pi)^4} \,\,
 \frac{k^\mu \,k^\nu}{(k^2 + i \,\varepsilon)^2 \,[(k-p)^2 + i \,\varepsilon]}
 \right. \nonumber \\
 &\,& \ \left. \hspace{6mm} - \,\,p \cdot n \,\int \,\frac{d^4 k}{(2 \,\pi)^4} \,\,
 \frac{k_\perp^2}{(k^2 + i \,\varepsilon)^2 \,[(k-p)^2 + i \,\varepsilon]} 
 \,\right\},
\end{eqnarray}
with $k_\perp^2 \equiv k_1^2 + k_2^2$. Evaluating its divergent part by the
dimensional regularization, we get
\begin{equation}
 T^A_{Gq} \ = \ \frac{\alpha_S}{2 \,\pi} \,\,
 \frac{3}{2} \,\,C_F \,\,\frac{1}{2 - \omega}.
\end{equation}

In this way, we arrive at the standardly-known answer for
$\Delta \gamma_{Gq}$ given by
\begin{equation}
 \Delta \gamma_{Gq} \ = \ \frac{\alpha_S}{2 \,\pi} \,\,\frac{3}{2} \,\,C_F .
\end{equation}

\vspace{5mm}
\begin{figure}[ht]
\begin{center}
\includegraphics[width=6.5cm]{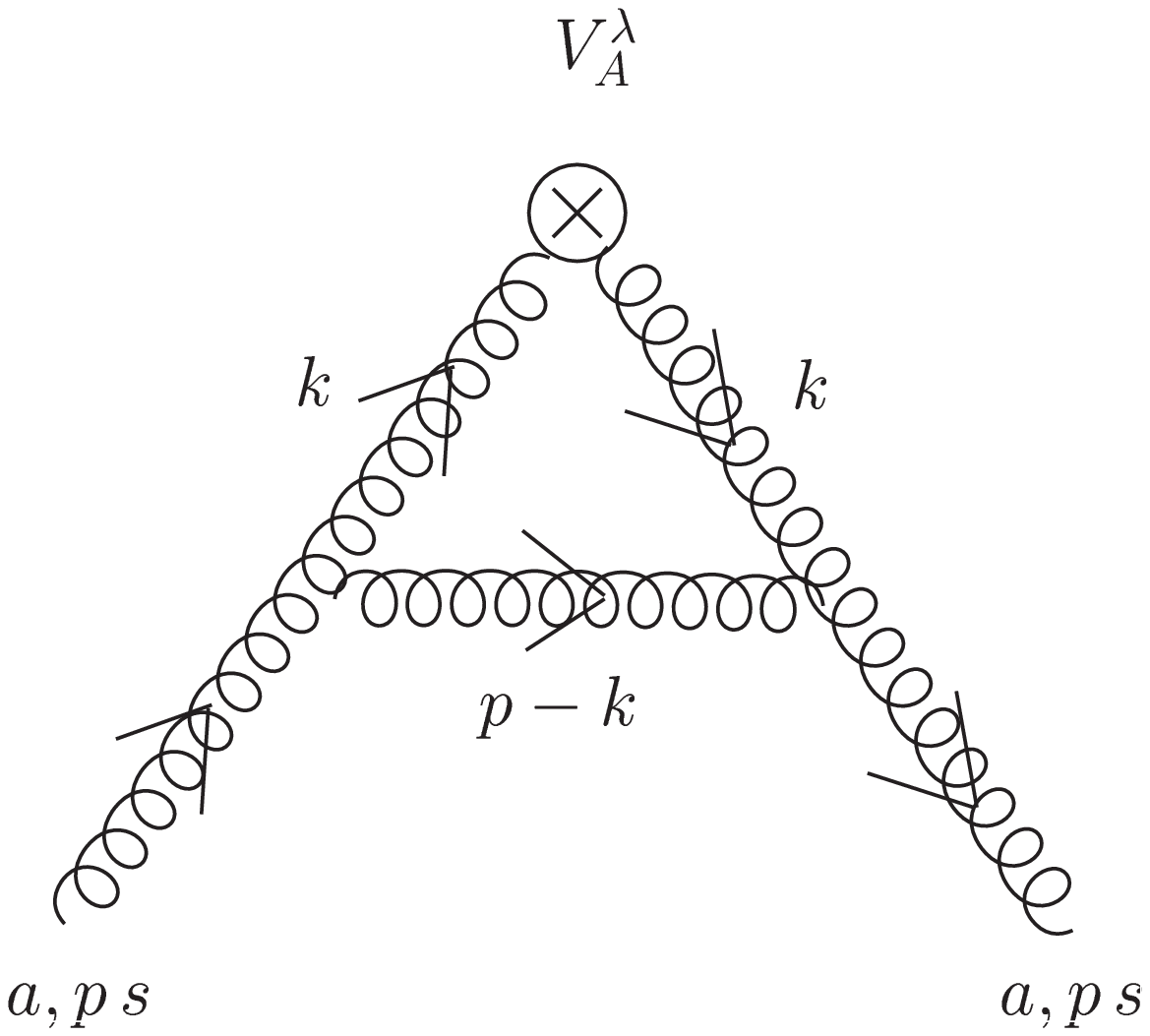}
\caption{The Feynman diagrams contributing to $\Delta \gamma_{GG}$.}
\label{Fig5_Vertex_GG}
\end{center} 
\end{figure}

Now we are in a position to investigate the most nontrivial part of
our analysis, i.e. the anomalous dimension $\Delta \gamma_{GG}$.
The contribution of the vertex $V_A$ is given by the Feynman diagram
illustrated in Fig.\ref{Fig5_Vertex_GG}(a). This gives
\begin{eqnarray}
 T^A_{GG} \ &=& \ \frac{1}{2 \,p \cdot n} \,\int \, \frac{d^4 k}{(2 \,\pi)^4} \,\,
 \epsilon^{\lambda *} (p s) \,\epsilon^\rho (p s) \nonumber \\
 &\,& \times \ g \,f^{a c^\prime e} \,\,
 [\,- \,(p + k)_\sigma \,g_{\lambda \nu^\prime} \ + \ 
 (2 \,k - p)_\lambda \,g_{\sigma \nu^\prime} \ + \ 
 (2 \,p - k)_{\nu^\prime} \,g_{\lambda \sigma} \,] \nonumber \\
 &\,& \times \ 2 \,i \,(k \cdot n) \,\,[\,g^{\mu 1} \,\delta^{\nu 2} (k) \ - \ 
 g^{\mu 2} \,\delta^{\nu 1} (k) \,] \,\,\delta^{bc} \nonumber \\
 &\,& \times \ g \,f^{a b^\prime d} \,\,
 [\,(p+k)_\tau \,g_{\rho \mu^\prime} \ + \ 
 (p - 2 \,k)_\rho \,g_{\mu^\prime \tau} \ + \ 
 (k - 2 \,p)_{\mu^\prime} \,g_{\rho \tau} \,] \nonumber \\
 &\,& \times \ \delta^{c c^\prime} \,D^{\nu \nu^\prime} (k) \,\,
 \delta^{b b^\prime} \,D^{\mu \mu^\prime} (k) \,\,
 \delta^{d e} \,D^{\tau \sigma} (p - k) .
\end{eqnarray}
This can be rewritten in the form : 
\begin{eqnarray}
 T^A_{GG} \ &=& \ + \,\,\frac{g^2 \,C_A}{p \cdot n} \,
 \int \,\frac{d^4 k}{(2 \,\pi)^4} \,\,
 \frac{k \cdot n}{(k^2 + i \,\varepsilon)^2 \,
 [(p-k)^2 + i \,\varepsilon]} \,\,
 \epsilon^{\lambda *} (p s) \,\,\epsilon^\rho (p s) \nonumber \\
 &\,& \times \ 
 [\,(p+k)_\sigma \,g_{\lambda \nu^\prime} \ + \ 
 (p - 2 \,k)_\lambda \,g_{\sigma \nu^\prime} \ + \ 
 (k - 2 \,p)_{\nu^\prime} \,g_{\lambda \sigma} \,] \nonumber \\
 &\,& \times \ 
 [\,(p+k)_\tau \,g_{\rho \mu^\prime} \ + \ 
 (p - 2 \,k)_\rho \,g_{\mu^\prime \tau} \ + \ 
 (k - 2 \,p)_{\mu^\prime} \,g_{\rho \tau} \,] \nonumber \\
 &\,& \times \ 
 (\,g^{\mu 1} \,g^{\nu 2} - g^{\mu 2} \,g^{\nu 1} \,) \,\,
 P_\nu{}^{\nu^\prime} (k) \,\,P_\mu{}^{\mu^\prime} (k) \,\,
 P^{\tau \sigma} (p - k),
\end{eqnarray}
where $C_A = f^{abc} \,f^{abc} = 3$ is the standard color factor.
After tedious but straightforward algebra, $T^A_{GG}$ can further be
rewritten in the form,
\begin{eqnarray}
 T^A_{GG} \ &=& \ 
 + \,\,C_A \,\,\frac{g^2}{p \cdot n} \,\int \,\frac{d^4 k}{(2 \,\pi)^4} \,
 \frac{k \cdot n}{(k^2 + i \,\varepsilon) \,[(k-p)^2 + i \,\varepsilon]} \,\,
 \epsilon^{\lambda *} (p s) \,\,\epsilon^\rho (p s) \nonumber \\
 &\,& \times \ \left\{ \, (\,g^{\mu^\prime 1} \,g^{\nu^\prime 2} \ - \ 
 g^{\mu^\prime 2} \,g^{\nu^\prime 1} \,) \right. \nonumber \\
 &\,& \ - \ (\,k^1 \,g^{\nu^\prime 2} - k^2 \,g^{\nu^\prime 1} \,) \,\,
 \frac{n^{\mu^\prime}}{k \cdot n} \ + \ 
 (\,k^1 \,g^{\mu^\prime 2} - k^2 \,g^{\mu^\prime 1} \,) \,\,
 \frac{n^{\nu^\prime}}{k \cdot n} \nonumber \\
 &\,& \ + \ \left. 
 (\,k^1 \,g^{\nu^\prime 2} - k^2 \,g^{\nu^\prime 1} \,) \,\,
 \frac{n^2 \,k^{\mu^\prime}}{(k \cdot n)^2} \ - \ 
 (\,k^1 \,g^{\mu^\prime 2} - k^2 \,g^{\mu^\prime 1} \,) \,\,
 \frac{n^2 \,k^{\nu^\prime}}{(k \cdot n)^2} \right\} \nonumber \\
 &\,& \times \ 
 [\,\epsilon^*_{\nu^\prime} \,(p+k)_\sigma \ - \ 
 2 \,\epsilon^* \cdot k \,\,g_{\sigma \nu^\prime} \ + \ 
 \epsilon^*_\sigma \,(k - 2 \,p)_{\nu^\prime} \,] \nonumber \\
 &\,& \times \ 
 [\,\epsilon_{\mu \prime} \,(p+k)_\tau \ - \ 
 2 \,\epsilon \cdot k \,\,g_{\tau \mu^\prime} \ + \ 
 \epsilon_\tau \,(k - 2 \,p)_{\mu^\prime} \,] \nonumber \\
 &\,& \times \ 
 \left[\, g^{\tau \sigma} \ - \ 
 \frac{(k-p)^\tau \,n^\sigma + n^\tau \,(k-p)^\sigma}{(k - p) \cdot n} \ + \ 
 \frac{n^2 \,(k-p)^\tau \,(k-p)^\sigma}{[(k-p) \cdot n]^2]} \,\right] .
\end{eqnarray}
We shall again calculate the three contributions from $P^{\tau \sigma} (k)$
separately. The part containing the tensor $g^{\tau \sigma}$ reduces to
\begin{equation}
 T^A_{GG} (g^{\tau \sigma}) \ = \ 
 T^{A_1}_{GG} (g^{\tau \sigma}) \ + \ T^{A_2}_{GG} (g^{\tau \sigma}) ,
\end{equation}
where
\begin{equation}
 T^{A_1}_{GG} (g^{\tau \sigma}) \ = \ - \,i \,\,C_A \,\,\frac{g^2}{p \cdot n} \,
 \int \,\frac{d^4 k}{(2 \,\pi)^4} \,\,
 \frac{k \cdot n \,(p+k)^2 - 8 \,p \cdot n \,k_\perp^2}
 {(k^2 + i \,\varepsilon) \,[(k-p)^2 + i \,\varepsilon]} ,
\end{equation}
and
\begin{equation}
 T^{A_2}_{GG} (g^{\tau \sigma}) \ = \ 2 \,i \,\, 
 C_A \,\,\frac{g^2}{p \cdot n} \,\,n^2 \,
 \int \,\frac{d^4 k}{(2 \,\pi)^4} \,\,
 \frac{k_\perp^2 \,(k^2 \,- \,3 \,p \cdot k)}
 {(k^2 + i \,\varepsilon)^2 \,[(k-p)^2 + i \,\varepsilon ] \,(k \cdot n)} .
\end{equation}
The 1st part, which does not contain the $1 \,/\,(k \cdot n)$ type
spurious singularity can be calculated in a standard manner, which gives
\begin{equation}
 T^{A_1}_{GG} (g^{\tau \sigma}) \ = \ + \,\,\frac{\alpha_S}{2 \,\pi} \,\,
 \frac{5}{2} \,\,C_A \,\,\frac{1}{2 - \omega} .
\end{equation}
The 2nd part can be evaluated by using the formulas : 
\begin{eqnarray}
 &\,& \mbox{div} \,\int \,d^{2 \omega} k \,\frac{k_\perp^2}
 {(k^2 + i \,\varepsilon) \,[(k-p)^2 + i \,\varepsilon ] \,[k \cdot n]}
 \ = \ - \,\frac{1}{D} \,\left(\,p \cdot n^* \ - \ 
 \frac{n^{* 2}}{n^* \cdot n + D} \,\,p \cdot n \,\right) \,\bar{I}, 
 \ \ \ \ \ \ \ \\
 &\,& \mbox{div} \,\int \,d^{2 \omega} k \,\frac{k^\mu \, k_\perp^2}
 {(k^2 + i \,\varepsilon) \,[(k-p)^2 + i \,\varepsilon ] \,[k \cdot n]}
 \ = \ - \,\frac{1}{2 \,D} \,\left(\,n^{* \mu} \ - \ 
 \frac{n^{* 2}}{n^* \cdot n + D} \,\,n^\mu \,\right) \,\bar{I}. \ \ \ \ 
\end{eqnarray}
The answer is given as
\begin{equation}
 T^{A_2}_{GG} (g^{\tau \sigma}) \ = \ - \,\frac{\alpha_S}{2 \,\pi} \,\,
 \frac{1}{2} \,\,C_A \,\,\frac{n^2}{D} \,\,\frac{1}{[p \cdot n]} \,\,
 \left( p \cdot n^* \ - \ \frac{n^{* 2}}{n^* \cdot n + D} \,\,
 p \cdot n \,\right) \,\,\frac{1}{2 - \omega}
\end{equation}
Collecting the two pieces, we thus arrive at
\begin{eqnarray}
 T^A_{GG} \,(g^{\tau \sigma}) \ &=& \ + \,\,\frac{\alpha_S}{2 \,\pi} \,\,
 \frac{5}{2} \,\,C_A \,\,\frac{1}{2 - \omega} \nonumber \\
 &\,& - \ \frac{\alpha_S}{2 \,\pi} \,\,\frac{1}{4} \,\,C_A \,\,\frac{n^2}{D} \,
 \frac{1}{[p \cdot n]} \,\,\left(\,p \cdot n^* \ - \ 
 \frac{n^{* 2}}{n^* \cdot n + D} \,\,p \cdot n \,\right) \,\,\frac{1}{2 - \omega} .
\end{eqnarray}
Next, we evaluate the term containing the spurious singularity of
$1 \,/\,(k-p) \cdot n$ in $P^{\tau \sigma} (k - p)$.
After lengthy algebra, we obtain
\begin{eqnarray}
 T^A_{GG} \,(1 \,/\,(k-p) \cdot n) \ &=& \ 
 - \,\,i \,\,C_A \,\,\frac{g^2}{p \cdot n}
 \nonumber \\
 &\,& \times \ \left\{\,- \,\,2 \,\int \,\frac{d^4 k}{(2,\pi)^4} \,\,
 \frac{k \cdot n}{(k^2 + i \,\varepsilon) \,[(k-p)^2 + i \,\varepsilon]} \right.
 \nonumber \\
 &\,& \ \ - \,\,4 \,p \cdot n \,\int \,\frac{d^4 k}{(2,\pi)^4} \,\,
 \frac{k \cdot n}{(k^2 + i \,\varepsilon) \,[(k-p)^2 + i \,\varepsilon] \,
 (k-p) \cdot n} \nonumber \\
 &\,& \ \ + \ 2 \,\,n^2 \,\int \,\frac{d^4 k}{(2,\pi)^4} \,\,
 \frac{k_\perp^2}{(k^2 + i \,\varepsilon) \,[(k-p)^2 + i \,\varepsilon] \,
 (k-p) \cdot n} \ \ \ \ \ \ \ \ \ \ \nonumber \\
 &\,& \left. \ \ - \,\,n^2 \,\int \,\frac{d^4 k}{(2,\pi)^4} \,\,
 \frac{k_\perp^2}{(k^2 + i \,\varepsilon) \,[(k-p)^2 + i \,\varepsilon] \,\,
 k \cdot n} \,\,\right\} .  
\end{eqnarray}
Using the known integral formulas,
\begin{eqnarray}
 &\,& \mbox{div} \int \,d^{2 \omega} k \,\,
 \frac{k \cdot n}{(k^2 + i \,\varepsilon) \,
 [(k-p)^2 + i \,\varepsilon] \,[(k-p) \cdot n]} \ = \ \bar{I}, \\
 &\,& \mbox{div} \int \,d^{2 \omega} k \,
 \,\frac{k_\perp^2}{(k^2 + i \,\varepsilon) \,
 [(k-p)^2 + i \,\varepsilon] \,[(k-p) \cdot n]} \ = \ \frac{1}{D} \,
 \left( p \cdot n^* \ - \ 
 \frac{n^{* 2}}{n^* \cdot n + D} \,\, p \cdot n \,\right) \,\bar{I}, 
 \ \ \ \ \ \ \ \ \\ 
 &\,& \mbox{div} \int \,d^{2 \omega} k \,\,
 \frac{k_\perp^2}{(k^2 + i \,\varepsilon) \,
 [(k-p)^2 + i \,\varepsilon] \,[k \cdot n]} \ = \ - \,\frac{1}{D} \,
 \left( p \cdot n^* \ - \ 
 \frac{n^{* 2}}{n^* \cdot n + D} \,\, p \cdot n \,\right) \,\bar{I},
\end{eqnarray}
we find that
\begin{eqnarray}
 T^A_{GG} \,(1 \,/\,(k-p) \cdot n) \ &=& \ \frac{\alpha_S}{2 \,\pi} \,
 \left( - \,\frac{5}{2} \,C_A \right) \,\frac{1}{2 - \omega} \nonumber \\
 &+& \ \frac{\alpha_S}{2 \,\pi} \,\,\frac{3}{2} \,\,C_A \,\,\frac{n^2}{D} \,\,
 \frac{1}{[p \cdot n]} \,\,\left( p \cdot n^* \ - \ 
 \frac{n^{* 2}}{n^* \cdot n + D} \,\,p \cdot n \,\right) \,
 \frac{1}{2 - \omega}. \ \ \ \ 
\end{eqnarray}

Finally, we evaluate the contribution of spurious double pole term
$1 \,/\,[(k-p) \cdot n]^2$ in $P^{\tau \sigma} (k-p)$.
After some algebra, we obtain
\begin{eqnarray}
 T^A_{GG} \,(1 \,/\,[(k-p) \cdot n]^2) \ = \ 
 - \,\,i \,\,C_A \,\,\frac{g^2}{p \cdot n} \,\,n^2 \,
 \int \,\frac{d^4 k}{(2 \,\pi)^4} \,\,
 \frac{k \cdot n}{[(k-p)^2 + i \,\varepsilon] \,[(k-p) \cdot n]^2} . \ \ \ \  
\end{eqnarray}
Now, by using the integral formula
\begin{equation}
 \mbox{div} \,\int \,d^{2 \omega} k \,\,
 \frac{k^\mu}{[(k-p)^2 + i \,\varepsilon ] \,
 [(k-p) \cdot n]^2} \ = \ p^\mu \,\,\frac{2}{D} \,\,
 \frac{n^{* 2}}{n^* \cdot n + D} \,\,\bar{I} ,
\end{equation}
we obtain
\begin{equation}
 T^A_{GG} \,(1 \,/\,[(k-p) \cdot n]^2) \ = \ 
 \frac{\alpha_S}{2 \,\pi} \,\,C_A \,\,\frac{1}{D} \,\,
 \frac{n^2 \,\,n^{* 2}}{n^* \cdot n + D} \,\,\frac{1}{2 - \omega} .
\end{equation}
Summing up the three contributions, we finally arrive at
\begin{equation}
 T^A_{GG} \ = \ \frac{\alpha_S}{2 \,\pi} \,\,C_A \,\,\frac{n^2}{D} \,\,
 \frac{p \cdot n^*}{[p \cdot n]} \,\,\frac{1}{2 - \omega} .
\end{equation} 
Again, it is instructive to consider several limiting cases.
In the light-cone limit with $n^0 = n^3 = 1 \,/\,\sqrt{2}$ and $n^2 = 0$,
one sees that $T^A_{GG}$ above vanishes. This is consistent with the direct
calculation in the light-cone gauge \cite{JTH96}.
On the other hand, in the temporal
limit with $n^0 = 1, \,n^3 = 0$ and $n^2 = 1$, the coefficient of
$1 \,/\,(2 - \omega)$ {\it diverges}, since $D \rightarrow 0$ in this limit.
However, for obtaining the anomalous dimension $\Delta \gamma_{GG}$,
we must also take account of the self-energy insertion to the external
gluon lines. The contribution of these diagrams
turn out to be (see, for instance, \cite{Gaigg88A},\cite{BassettoBook})
\begin{eqnarray}
 T^{Self}_{GG} &=& \frac{\alpha_S}{2 \,\pi} \,
 \left( \frac{11}{6} \,C_A \ - \ \frac{1}{3} \,n_f \right) \,\,
 \frac{1}{2 - \omega} \ - \ 
 \frac{\alpha_S}{2 \,\pi} \,\,C_A \,\,\frac{n^2}{D} \,\,
 \frac{p \cdot n^*}{[p \cdot n]} \,\,\frac{1}{2 - \omega} .
\end{eqnarray}
One finds that the dangerous terms in $T_{GG}$ and $T^{Self}_{GG}$ cancel
exactly, thereby being led to
\begin{eqnarray}
 T^A_{GG} \ + \ T^{Self}_{GG} \ = \ \frac{\alpha_S}{2 \,\pi} \,
 \left( \frac{11}{6} \,C_A \ - \ \frac{1}{3} \,n_f \right) \,\,
 \frac{1}{2 - \omega} ,
\end{eqnarray}
which gives
\begin{equation}
 \Delta \gamma_{GG} \ = \ \frac{\alpha_S}{2 \,\pi} \,
 \left( \frac{11}{6} \,C_A \ - \ \frac{1}{3} \,n_f \right)
\end{equation}
In this way, we have succeeded in reproducing the well-known answer completely
independently of the choice of the 4-vector $n^\mu$, which is interpreted to
characterize the Lorentz frame in which the gauge-fixing condition is imposed.
The flexibility of our treatment on the choice of the 4-vector $n^\mu$
enables us to handle several interesting cases in a unified way with the
help of the generalized $n^*_\mu$-prescription. They include the temporal
gauge limit with $n^2 = 1$, the light-cone gauge limit with $n^2 = 0$,
and also the spatial axial-gauge limit with $n^2 = -1$, etc.
We have shown that the temporal gauge limit should be treated with special
care, because singular terms appear in the course of manipulation.
Nevertheless, after summing up all the relevant contributions,
dangerous singular terms cancel among themselves and the final answer
is shown to be the same in all the cases.
As we have shown, since these three different gauges belonging to the
general axial gauge can also be connected with different choices of path in
the geometric formulation, what we have shown is also interpreted as
the path-independence of the longitudinal gluon spin, although within
a restricted class of path choices. 
Undoubtedly, this is a gauge-invariance in a traditional sense.


Before ending this section, we make several supplementary remarks on the
significance of our finding above. In the previous paper \cite{Wakamatsu11A},
we gave a formal proof that the quark and
gluon dynamical OAMs appearing in our nucleon spin decomposition (I) can be related
to the difference between the 2nd moment of the unpolarized GPDs and
the 1st moment of the longitudinally polarized PDFs as
\begin{eqnarray}
 L_q \ &=& \ \langle p s \,|\,n_\lambda \,M^{\lambda 1 2}_{q-OAM} \,|\, p s \rangle
 \,/\, (n \cdot p) \nonumber \\
 &=& \frac{1}{2} \,\int \,x \,[\,H^q (x,0,0) \ + \ E^q (x,0,0) \,] \,\,dx \ - \ 
 \frac{1}{2} \,\int \,\Delta q(x) \,\,dx , \label{Eq:Lq_sum_rule} 
\end{eqnarray}
and
\begin{eqnarray}
 L_G \ &=& \ \langle p s \,|\, n_\lambda \,M^{\lambda 1 2}_{G-OAM} \,|\, p s \rangle
 \,/\, (n \cdot p) \nonumber \\
 &=& \frac{1}{2} \,\int \,x \,[\,H^g (x,0,0) \ + \ E^g (x,0,0) \,] \,\,dx \ - \ 
 \int \,\Delta g(x) \,\,dx. \label{Eq:LG_sum_rule} 
\end{eqnarray}
As is widely-known, (\ref{Eq:Lq_sum_rule}) is first derived by Ji.
The relation (\ref{Eq:LG_sum_rule}) was also written down by Ji,
but as an ad hoc definition of the gluon orbital angular momentum.
This is because his viewpoint is that the decomposition (\ref{Eq:LG_sum_rule})
is not a {\it truely} gauge invariant one.
It would be instructive to reconsider these relations in the context of
gauge-invariant-extension approach using gauge link or Wilson line.
It is widely accepted that the gauge-invariant definitions of the GPDs as well as
the polarized PDFs necessarily require the gauge link connecting two different
space-time point. However, the quantities appearing in the r.h.s. of the above
relations are not GPDs and PDFs themselves but their lower moments.
In fact, the above relations can also be expressed as \cite{Ji98}
\begin{eqnarray}
 L_q \ &=& \ \frac{1}{2} \,[\,A^q_{20} (0) \ + \ B^q_{20} (0) \,] \ - \ 
 \frac{1}{2} \,a^q (0), \\
 L_G \ &=& \ \frac{1}{2} \,[\,A^G_{20} (0) \ + \ B^G_{20} (0) \,] \ - \ 
 a^G (0) .
\end{eqnarray}
Here, $A^q_{20} (0), B^q_{20} (0), A^G_{20} (0)$ and $B^G_{20} (0)$ are
the forward limit ($t \rightarrow 0$) of the gravitational form factors
$A^q_{20} (t), B^q_{20} (t), A^G_{20} (t)$ and $B^G_{20} (t)$, while
$a^q (0)$ and $a^G (0)$ are the axial charges of quarks and gluons
corresponding to the forward limits of axial form factors $a^q (t)$ and $a^G (t)$.
(We recall that the quark and gluon axial charges are identified with
the quark and gluon intrinsic spins in the gauge-invariant $\overline{MS}$
regularization scheme, i.e. $a^q (0) = \Delta \Sigma$ and
$a^G (0) = \Delta G$.)
Note that, to extract the form factors, deep-inelastic-scattering measurements
are not mandatory.
For example, the gravitational form factors can in principle be extracted from
graviton-nucleon elastic scattering just as the electromagnetic form factors can
be extracted from electron-nucleon elastic scatterings, 
even though this is just a Gedanken experiment.
This means that, at least for these quantities,
i.e. for the form factors, we do not need to stick to such an idea that the path of
gauge-link has a physical meaning as claimed in gauge-invariant-extension approach.
In fact, we have explicitly demonstrated the path-independence of the evolution
matrix for the quark and gluon spins, although within a restricted class of choices
called the general axial gauges specified by the direction of the infinitely
long path.
This indicates that at least the above relations (\ref{Eq:Lq_sum_rule}) and
(\ref{Eq:LG_sum_rule}) are not affected by
continuous deformation of the path of Wilson lines used in the definitions
of the GPDs and the polarized PDFs.

Also worth remembering is the following well-known but sometimes
unregarded fact. Why does not one need to pay
much attention to the notion of path-dependence of the Wilson-line
in the case of the standard collinear PDFs ?
For clarity, let us first consider the simplest leading-twist PDF, i.e.
the unpolarized PDF. The modern way of defining the
unpolarized quark distribution function is to use the bilinear quark
operator with light-cone separation. The non-local Wilson line is
necessary here to ensure the gauge-invariance of the bilocal quark
operator. However, this definition of PDF is known to be completely
equivalent to the one based on the operator-product-expansion (OPE).
That is, the bilinear and
bilocal quark field with Wilson line is equivalent to the infinite tower
of local and gauge-invariant operators with higher covariant derivative.
Since these infinite tower of gauge-invariant operators are just
local operators although with higher derivatives, they are free from the
notion of path, i.e. they are independent of particular direction in space
and time. The situation is simply the same also for other PDFs.
Within the framework of OPE, the gluon distribution can also be
defined in terms of infinite towers of local and gauge-invariant operators.
Namely, within the framework of the OPE, they can be defined without calling
for the notion of paths. (Only one important exception is the gluon spin
operator corresponding to the 1st moment of the longitudinally polarized
gluon distribution functions discussed in the present paper.
The long known worrying fact was that, as long as one stick to the locality,
there is no twist two spin-one gauge-invariant gluon operator.)
By this reason, the notion of path-dependence of the Wilson-line has seldom
been made an issue of at least in the case of collinear PDFs.
The same can be said also for the GPDs appearing in the sum rules
(\ref{Eq:Lq_sum_rule}) and (\ref{Eq:LG_sum_rule}).


Unfortunately, such a simplification cannot be expected for the
transverse-momentum dependent PDFs or more general Wigner distributions. 
This is the reason why the status for another gauge-invariant
decomposition (II) is still in unclarified status.
In fact, a very interesting relation between the OAMs and Wigner
distributions was first suggested by Lorc\'{e} and Pasquini \cite{LorcePasquini11}.
However, the gauge-invariant definition of Wigner distribution requires gauge
link or Wilson line, which is generally path-dependent.
Hatta showed that the LC-like path choice gives ``canonical'' OAM \cite{Hatta12}.
On the other hand, Ji, Xiong, and Yuan argued that the straight path connecting
the relevant two space-time points gives ``dynamical" OAM \cite{JXY12A}. 
Assuming that both are correct, one might be lead to two possible scenarios.
The 1st possibility is that, because there are infinitely many paths
connecting the two relevant space-time points appearing in the gauge-invariant
definition of Wigner distribution, there are infinitely many
Wigner distributions and consequently infinitely many quark and gluon OAMs.
The 2nd possibility is that the Wigner distributions with infinitely many
paths of gauge-link are classified into some discrete pieces or
equivalent classes, which cannot be continuously deformable into each other.
The recent consideration by Burkardt may be thought of as an indication of
this 2nd possibility \cite{Burkardt12}.
At any rate, it would be fair to say that, at least up to now,  we do not
have any convincing answer to the question of the real observability of the
nucleon spin decomposition (II).

\section{conclusion}
We have investigated the uniqueness or non-uniqueness problem of the
decomposition of the gluon field into the physical and pure-gauge
components, which is the basis of the recently proposed two physically
inequivalent gauge-invariant decomposition of the nucleon spin.
It was emphasized that, the physical motivation of this decomposition is
the familiar transverse-longitudinal decomposition in QED, which is known
to be unique once the Lorentz-frame of reference is fixed.
In the case of nonabelian gauge theory, this transverse-longitudinal
decomposition becomes a little more nontrivial even in the noncovariant
treatment. In fact, the past researches reveal the fact that the transverse
component of the nonablelian gauge field can be expressed only in a
perturbation series in the gauge coupling constant.
Nevertheless, it is very important to recognize the fact that to project
out the physical component of the gauge field is essentially equivalent
to the process of gauge-fixing. In fact, in the geometrical formulation
of the nonabelian gauge theory, a closed form of the physical component
of the gauge field is known, although it requires the non-local
Wilson line, depending on a path in the 4-dimensional space and
time. It is also known that a choice of path is inseparably connected
with a choice of gauge. 
An especially useful choice for our purpose of defining a gauge-invariant
gluon spin operator is an infinitely long straight-line path connecting
the space-time point of the gauge field and the space-time infinity, the direction of
which is characterized by a constant 4-vector. This particular choice
of path is known to be equivalent to taking the so-called general axial gauge,
which contains three popular gauges, i.e. the temporal, the light-cone,
and the spatial axial gauges. Based on this general axial gauge, characterized
by the constant 4-vector $n^\mu$, we have calculated the 1-loop anomalous
dimension matrix for the quark and gluon longitudinal spins in the nucleon.
We then find that the final answer is exactly the same independently the
choice of $n^\mu$, which amounts to proving the gauge-independence and
the path-independence simultaneously.
After all, what we have explicitly shown is only the perturbative gauge-
and path-independence of the gluon spin. Nevertheless, our general argument
offers strong counter-evidence to the idea that there are infinitely many
decompositions of the nucleon spin. It also give a support to our claim
that the total angular momentum of the gluon can be gauge-invariantly
decomposed into the orbital and intrinsic spin parts as long as the
longitudinal spin sum rule of the nucleon is concerned.
This means that the dynamical OAMs of quarks and gluons appearing in our
decomposition (I) can be thought of as genuine observables, in the sense
that there is no contradiction between this decomposition and the general
gauge principle of physics. 

On the other hand, the observability of the OAM appearing in the decomposition (II),
i.e. the generalized ``canonical" OAM, is not completely clear yet.
This is because, although the relation between the ``canonical" OAM and a
Wigner distribution is suggested, its path-dependence or path-independence
should be clarified more convincingly.
Moreover, once quantum loop effects are included, the very existence of TMDs
as well as Wigner distributions satisfying gauge-invariance and factorization
(or universality) at the same time is under debate. (See \cite{BookCollins},
and references therein.)
Is process-independent extraction of ``canonical'' OAM possible ?
This is still a challenging open question.

\vspace{3mm}
\begin{acknowledgments}
The author greatly acknowledge many stimulating discussions with Cedric Lorc\'{e},
although our opinions still differ in several essential points.
He also greatly appreciates useful discussion with Takahiro Kubota.
\end{acknowledgments}

\vspace{10mm}

\end{document}